\documentclass[superscriptaddress,twocolumn,amsmath,amssymb,aps,prb]{revtex4-1}
\usepackage[dvips]{graphicx}
\usepackage{dcolumn}
\usepackage{bm}

\usepackage[dvips]{graphicx}
\usepackage{dcolumn}
\usepackage{color}
\usepackage{bm}
\usepackage[colorlinks=true,citecolor=blue,linkcolor=blue,pdfhighlight =/O]{hyperref}

\usepackage{epstopdf}

\renewcommand{\Re}{\mathop{{\rm Re}}\nolimits}
\newcommand{\St}{\mathop{{\rm St}}\nolimits}
\newcommand{\ep}{\epsilon}
\newcommand{\mF}{\mathcal{F}}
\newcommand{\mT}{\mathcal{T}}

\begin{document}

\title{Recovery of a SINIS turnstile accuracy in a strongly non-equilibrium regime}

\author{I. M. Khaymovich}
\affiliation{Laboratoire de Physique et Mod\'elisation des Milieux Condens\'es, Universit\'e de Grenoble Alpes and CNRS, 25 rue des Martyrs, 38042 Grenoble, France}
\affiliation{Nanosciences Foundation, 23 rue des Martyrs, 38000 Grenoble, France}
\affiliation{Institute for Physics of Microstructures, Russian Academy of Sciences, 603950 Nizhny Novgorod, GSP-105, Russia }
%

\author{D. M. Basko}
\affiliation{Laboratoire de Physique et Mod\'elisation des Milieux Condens\'es, Universit\'e de Grenoble Alpes and CNRS, 25 rue des Martyrs, 38042 Grenoble, France}

\begin{abstract}
We perform a theoretical study of non-equilibrium effects
in charge transport through a hybrid single-electron transistor
based on a small normal metal (N) island with the gate-controlled number of electrons,
tunnel-coupled to voltage-biased superconducting (S) electrodes (SINIS).
Focusing on the turnstile mode of the transistor operation with the gate voltage driven periodically,
and electrons on the island being out of equilibrium,
we find that the current quantization accuracy is a non-monotonic function of the relaxation rate $\Gamma_{\mF}$ of the distribution function $\mF(\ep)$ on the island due to tunneling,
as compared to the drive frequency $f$, electron-electron $1/\tau_{ee}$ and electron-phonon $1/\tau_{eph}$ relaxation rates.
Surprisingly, in the strongly non-equilibrium regime, $f\gg \Gamma_{\mF}\gg\tau_{ee}^{-1},\tau_{eph}^{-1}$, the turnstile current plateau is recovered, similarly to the ideal equilibrium regime, $\tau_{eph}^{-1}\gg \Gamma_{\mF}$.
The plateau is destroyed in the quasiequilibrium regime when the electron-electron relaxation is faster than tunneling.
\end{abstract}




\date{\today}
\maketitle

\section{Introduction}
Nowadays, hybrid superconducting systems play an important role in several domains of physics and technology such as electronic refrigeration \cite{Giazotto2006}, metrology \cite{Pekola_RMP_2013} etc.
One aspect, widely addressed in the literature, is the overheating of the superconducting (S) parts of hybrid junctions as they can be easily driven out of equilibrium under typical operating conditions (see, e.g., Ref.~\onlinecite{Knowles2012}).
When overheated, superconducting parts contain many hot quasiparticles (QPs) which compromise the performance of mesoscopic devices.\cite{Footnote_QP_poisoning}
To overcome the problem of hot QPs, quite a number of efforts have been made to reduce these overheating effects in S by putting different types of quasiparticle traps \cite{Peltonen2011, Nsanzineza2014, Wang2014, Vool2014, Woerkom2015, Taupin2016_NatComm} and by cooling S-parts directly.\cite{Blamire1991,Heslinga1993,Goldie1990,Nguyen2013}
Despite some pessimistic theoretical predictions of the residual QP densities,\cite{Bespalov_Nazarov_2016} these efforts have been rather successful as the quasiparticle densities were indeed remarkably reduced there.\cite{Pekola_PRL_2010,Knowles2012, Saira2012, Riste2013, Woerkom2015}

Besides the problem of hot QPs in the superconductors, which is partly solved, there is a question of electron overheating in the normal metal (N) parts of mesoscopic devices. The relaxation rates in N parts are significantly faster than in S; nevertheless, there are both theoretical and experimental studies of the non-equilibrium distributions in N wires,\cite{Pothier1997} in single-electron transistors,\cite{Nazarov_Heikkila2010} and in N parts of NIS coolers \cite{Pekola_Heikkila_Non-eq_f(E)-1,Pekola_Heikkila_Non-eq_f(E)-2} where `I' stands for insulator.
Under experimentally achievable conditions, the electronic energy distribution $\mF(\ep)$ in the normal metal can have non-Fermi-Dirac form, which leads to measurable consequences.

\begin{figure}[h]
\center{
\includegraphics[width=0.25\textwidth]{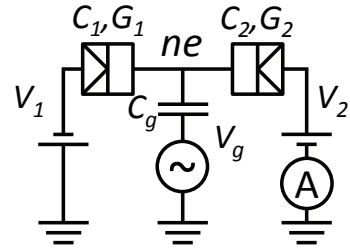}
}
\caption{
Schematics of a hybrid SINIS SET used in turnstile experiments.
}\label{Fig1:Schematics}
\end{figure}

In this paper, we address the problem of non-equilibrium electronic distributions in a similar type of device, namely, in a hybrid superconducting single-electron transistor (SET) consisting of a small metallic normal island, sandwiched between two superconducting electrodes. Such configuration, usually called SINIS, is shown schematically in Fig.~\ref{Fig1:Schematics}.
Under equilibrium conditions, a large Coulomb energy
(compared to the bath temperature~$T$) prevents an extra electron from tunneling into the island, which enables one to control the electron number $n$ of the island by applying a voltage to a nearby gate electrode.

One way to drive a SINIS SET out of equilibrium is to apply a periodic driving gate voltage in the so-called turnstile regime,\cite{SINIS_turnstile_Nat2008} which has potential applications in quantum metrology.\cite{Pekola_RMP_2013, Kashcheyevs_RPP2015} Then, one measures the charge current $\langle I\rangle$ through the device, averaged over the drive period.
%
\begin{figure}[h]
\center{
\includegraphics[width=0.45\textwidth]{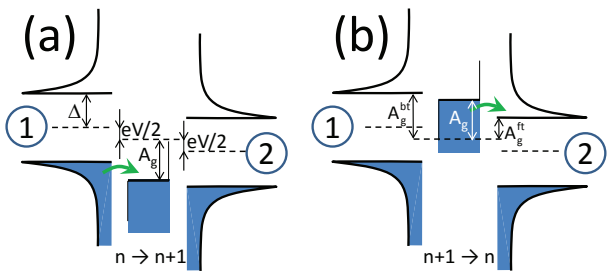}
}
\caption{(Color online) An illustration of the SINIS turnstile operation in the equilibrium regime.
Panels show energy diagrams of quasiparticle distributions in superconducting electrodes with a gap $\Delta$ and in the normal island for the two stages of the turnstile operation.
The two positions $\mu=\mp A_g$ of the island chemical potential,
shown by solid black lines, correspond to (a) injection and (b) ejection stages, respectively.
The filled (empty) states are shown by blue (white) color, while the electrodes are numbered by $1$ and $2$ in the circles.
Both in the island and in electrodes, the electronic distributions are zero-temperature Fermi distributions.
The electrode chemical potentials are shifted by a bias voltage $V$ and are shown by horizontal dashed lines.
The main tunneling process of injection (ejection) is shown by the green arrow.
}\label{Fig2:SINIS_turnstile}
\end{figure}
A hybrid superconducting turnstile was realized for the first time in Ref.~\onlinecite{SINIS_turnstile_Nat2008}. It is a voltage-biased hybrid SINIS SET working as schematically shown in Fig.~\ref{Fig2:SINIS_turnstile}. By applying a periodic voltage to the gate electrode with a driving frequency $f$, one can change the chemical potential $\mu$ of the normal island (see the two positions of $\mu(t)$ in panels (a) and (b) of the figure). Due to the Coulomb interaction, in a certain range of the gate voltage amplitudes only two charge states of the island are available. Therefore, when $\mu$ is aligned with the lower quasiparticle branch below the gap of the left electrode, one electron is loaded to the island from the left electrode, see panel (a), while when $\mu$ is above the gap of the right electrode, an electron tunnels out from the island to the right electrode, panel (b). This results in a quantized average current $\langle I\rangle = e f$, with $e>0$ being the elementary charge. In the present paper, we study how the non-equilibrium electron distribution on the island affects the device operation.

The paper is organized as follows.
In Sec.~\ref{Sec:model} we formulate the theoretical model and write the main equations describing the electron dynamics in the turnstile.
Sec.~\ref{Sec:Qualitative_Discussion} qualitatively explains the main ideas and the results of the paper.
In Sec.~\ref{Sec:Fast_drive} we study the most interesting regime of the turnstile operation, when the drive is not too slow, and estimate the relevant experimental parameters.
In Sec.~\ref{App:slow_drive}, we show that in the opposite case (slow drive) the turnstile errors are very large, so this regime is not interesting.
In the last section, we summarize our results.

\section{The model}\label{Sec:model}
We assume the single-electron energy levels on the normal
island to be randomly distributed with the mean level spacing
$\delta=[\nu_\mathcal{N}(0)\mathcal{V}]^{-1}$, where $\nu_\mathcal{N}(0)$ and
$\mathcal{V}$ are the density of states (DOS) per spin projection
and the island volume, respectively.
Depending on the island size, $\delta$ can be large or small compared to other energy scales of the problem.
Turnstile devices based both on large\cite{SINIS_turnstile_Nat2008} and small\cite{vanZanten2016} islands have been realized.
Here we focus on the limit of large islands (small $\delta$).
We also restrict ourselves to
the situation without any external magnetic field, so each level
is doubly degenerate with respect to the spin. Tunnel coupling
to the electrodes broadens the single-particle levels.
Characterizing each tunnel junction $j=1,2$ by the conductance
$G_j$ it has when the electrode is in the normal state (see Fig.~\ref{Fig1:Schematics}), we can
relate the average level broadening $\gamma_1+\gamma_2$
($2\gamma_j$ being the average escape rate from a single level
on the island to the $j$th electrode in the normal state) to
the conductances as $G_j=2e^2\gamma_j/\delta$, where the factor
of~2 keeps track of the spin degeneracy
and we set the Planck constant $\hbar=1$.
We assume the tunnel coupling to be weak, $\gamma_j\ll\delta$, to ensure strong Coulomb blockade.\cite{Aleiner2002}

Besides the single-particle contribution to the
electronic energy of the island, we include the Coulomb
electrostatic contribution. It is determined by the total
charge $Q=-ne$ on the island, where $n$ is the total number
of the excess electrons there.
In addition, one can control the
electrostatic energy by applying a voltage $V_g$ to the
gate electrode (see Fig.~\ref{Fig1:Schematics}). We write the Coulomb
energy as $E_n=E_C(n^2-2nn_g)$, where the charging energy
$E_C=e^2/2C_\Sigma$ is governed by the island total capacitance
$C_\Sigma=C_1+C_2+C_g$ being the sum of the capacitances to each
of the electrodes $C_j$, $j=1,2$, as well as of the capacitance
to the gate electrode $C_g$, and $n_g\equiv{C}_gV_g/e$. 
Here we neglect the effects of overheating of S leads due to the success of the QP reduction mentioned in the introduction and
assume the electrodes to be in thermal equilibrium at temperature~$T\ll\Delta$ small compared to
the superconducting gap $\Delta$,
and at different chemical potentials $-eV_j$ determined by the
applied constant bias voltage $V$.
For simplicity we consider a symmetric device with
$C_1=C_2$, $G_1=G_2$, $\gamma_1=\gamma_2=\gamma$, and
$-V_1=V_2=V/2>0$.

The electronic state of the island is specified by the total electron
number~$n$ (always an integer), and by the occupations of
all single-electron levels. In the statistical description,
one works with the probability $p_n$ to have $n$~excess
electrons, and with the conditional occupation probability
$\mF_n(\ep)$ of a single-particle state with
energy~$\ep$ and a given spin, provided that the island
has exactly $n$~excess electrons.
They are subject to the constraints $\sum_np_n=1$ and
\begin{gather}\label{F_n_fixed_mu}
\frac{2}{\delta}\int[\mF_n(\ep)-\theta(-\ep)]\,d\ep=n \ .
\end{gather}
As any fermionic distribution function,
$\mF_n(\ep)$ has the limiting values
$\mF_n(\ep\to\pm\infty)=0,1$.

We assume the island to be in the strong Coulomb blockade
regime, which occurs when $T\ll E_C$ and $\gamma\ll\delta$.
Then, if the gate voltage varies within the interval $0<n_g<1$,
$p_n$~is dominated by at most two values of~$n$ close to the
minimum of $E_n$, $n=0,1$.
As mentioned before we assume the island to be large enough, so that
$E_C,e V,\Delta\gg\delta$.
Then, many single-particle levels can participate in the transport, so a change of~$n$ by~1
translates into a very small change of the occupation
probability of each individual level and into a change of
the chemical potential by a small amount $\sim\delta$.
As a result, one can
neglect the difference between the distribution functions
$\mF_0(\ep)$, $\mF_1(\ep)$, and their
average $\mF(\ep)\equiv
p_0\mF_0(\ep)+p_1\mF_1(\ep)$ at any given~$\ep$ (see Appendix~\ref{App:check_f_0=f_1}).
Under the above conditions, one can write a closed set
of rate equations for $p_n$ and $\mF(\ep)$, as it
was done in Refs.~%
\onlinecite{Averin-Korotkov_JETP90,Averin-Korotkov_JLTP90}:
\begin{subequations}\label{rate_eq}
\begin{align}
\label{rate_eq_pn}
&\dot{p}_1 = -\dot{p}_0 = \sum_j\frac{2\gamma}\delta
\left[p_0\,w_+(U_j) - p_1\,w_-(U_j)\right],\\
&w_+(U_j)\equiv\int d\ep\,n_S(\ep)\,\mF_T(\ep)
\left[1-\mF(\ep-U_j)\right],
\label{w+=}\\
&w_-(U_j)\equiv\int d\ep\,n_S(\ep)\,\left[1-\mF_T(\ep)\right]
\mF(\ep-U_j)\ ,
\label{w-=}
\end{align}
\end{subequations}
\begin{equation}\begin{split}\label{kinetic_eq}
\dot{\mF}(\ep,t) ={}&{} \gamma{p}_0\left[1-\mF(\ep,t)\right]
\sum_j n_S(\ep+U_j)\,\mF_T(\ep+U_j)-{}\\
{}&{}-\gamma{p}_1\mF(\ep,t)
\sum_j n_S(\ep+U_j)\left[1-\mF_T(\ep+U_j)\right]+{}\\
{}&{}+ \St[\mF](\ep) \ ,
\end{split}\end{equation}
where the dot denotes the time derivative. Here we introduced
$n_S(\ep)=\left|\Re\left(\ep/\sqrt{\ep^2-\Delta^2}\right)\right|$,
the quasiparticle DOS in the superconducting
electrodes normalized to its normal-state value,
 and $U_j\equiv\mu+eV_j$
is the change of electrostatic energy due to electron tunneling
into the island from the $j$th electrode with
$\mu\equiv E_{n+1}-E_n=2 E_C(n+1/2-n_g)$ playing the role of
the chemical potential of the island.
$\mF_T(\ep)\equiv{1}/(1+e^{\ep/T})$ is the Fermi-Dirac
distribution with the bath temperature $T$.

The last term in Eq.~\eqref{kinetic_eq},
$\St[\mF](\ep)$,
is the collision integral which describes relaxation of the electronic
energy distribution towards thermal equilibrium.
We consider two relaxation mechanisms, electron-electron and
electron-phonon collisions on the island (see, e.~g.,
Ref.~\onlinecite{Giazotto2006} for a review),
\begin{subequations}\label{St=}
\begin{equation}
\St[\mF](\ep) = \St_{ee}[\mF](\ep)+\St_{eph}[\mF](\ep).
\end{equation}
We describe both in the $\tau$-approximation:
\begin{align}
&\St_{ee}[\mF](\ep)=\frac{\mF_{T_e}(\ep)-\mF(\ep)}{\tau_{ee}},
\label{Stee=}\\
&\St_{eph}[\mF](\ep)=\frac{\mF_{T}(\ep)-\mF(\ep)}{\tau_{eph}}.
\label{Steph=}
\end{align}
\end{subequations}
Here $\mF_{T_e}(\ep)=1/(1+e^{\ep/T_e})$ is the Fermi-Dirac
distribution with a certain temperature~$T_e$, chosen so that
the total energy of the limiting equilibrium distribution
$\mF_{T_e}(\ep)$ matches that in the distribution $\mF(\ep)$:
\begin{equation}\label{T_e}
\frac{2}\delta\int \ep\left[\mF_{T_e}(\ep)-\theta(-\ep)\right]d\ep
= \frac{2}\delta\int \ep\left[\mF(\ep)-\theta(-\ep)\right]d\ep \ ,
\end{equation}
where the left-hand side is equal to $\pi^2T_e^2/(3\delta)$. This ensures that
$\St_{ee}[\mF](\ep)$ conserves the total electronic energy.\cite{Footnote_mu_in_St}
The electron-phonon processes do not conserve the electronic
energy and drive the electronic distribution towards
$\mF_T(\ep)$.

The two relaxation times $\tau_{ee}$ and $\tau_{eph}$ may
depend on electronic temperature $T_e$, 
but not on the energy~$\ep$ (such
dependence would lead to violation of particle and energy
conservation). For $\tau_{ee}$, we assume the temperature
dependence corresponding to the zero-dimensional
limit:\cite{Sivan1994,Blanter1996}
\begin{equation}\label{SIA=}
\frac{1}{\tau_{ee}}=\delta\,\frac{T_{e}^2}{E_{\rm Th}^2},
\end{equation}
where $E_{\rm Th}$ is the Thouless energy of the island,
defined by the order of magnitude as the inverse of the
time required for an electron to travel across the island,
thus randomizing its motion due to scattering off impurities
or the dot boundaries, $E_{\rm Th}\sim\min\{v_F/L,D/L^2\}$,
where $L\sim \mathcal{V}^{1/3}$~is the typical island size, $v_F$ is the Fermi velocity,
and $D$~is the electron diffusion coefficient.
In Eq.~(\ref{SIA=}), we omitted the numerical prefactor
which is determined by the island shape.

As for the temperature dependence of~$\tau_{eph}$, it can be
conveniently determined by considering the total power~$W$
transferred to phonons from electrons whose distribution
$\mF_{T_e}(\ep)$ is thermal, but with a temperature~$T_e$
different from the phonon temperature~$T$. It is related
to the electron-phonon collision integral as
\begin{equation}\label{W=St}
W=\frac{2}\delta\int\ep\St_{eph}[\mF_{T_e}](\ep)\,d\ep.
\end{equation}
For this cooling power, a variety of expressions is available
in the literature, which were derived microscopically in
different regimes determined by the system dimensionality,
the shape of the Fermi surface, the relation between the
phonon wavelength, island size, and the electron mean free
path.%
\cite{Wellstood1994,Sergeev2000,Yudson2003,Basko2005,%
Sergeev2005,Prunnila2005}
All these expressions can be represented in the form
\begin{equation}\label{coolingpower=}
W=\frac{\pi^2}{3\delta}\,\frac{T_e^\alpha-T^\alpha}{T_{eph}^{\alpha-3}},
\end{equation}
with some power $\alpha$ between 4 and~6, and with $T_{eph}$
parameterizing the electron-phonon coupling strength.
The proportionality of the cooling power to the island
volume $\mathcal{V}$ is ensured by the factor
$1/\delta=\nu_\mathcal{N}(0)\mathcal{V}$ in
Eq.~(\ref{coolingpower=}).
Strictly speaking, it is impossible to obtain an expression
of the form~(\ref{coolingpower=}) from Eq.~(\ref{W=St}) in the
$\tau$~approximation~(\ref{Steph=}) with an energy-independent
$\tau_{eph}$. Still, if we use the following model dependence:
\begin{equation}\label{taueph=}
\frac{1}{\tau_{eph}}=
\frac{\max\{T^{\alpha-2},T_e^{\alpha-2}\}}{T_{eph}^{\alpha-3}},
\end{equation}
Eq.~(\ref{coolingpower=}) matches expression~(\ref{W=St}) when
$T_e$ and $T$ are strongly different. It gives a wrong numerical
coefficient in~$W$ when $T_e\approx{T}$, but the
$\tau$~approximation is valid only qualitatively anyway.
In the subsequent numerical calculations, we take $\alpha=5$,
so that Eq.~\eqref{coolingpower=} can be equivalently represented in the form $W=\Sigma \mathcal{V}\,(T_e^5-T^5)$,
where $\Sigma$ is a material constant.\cite{Giazotto2006}

The specific observable we are interested in, is the steady-state period-averaged current $\langle I\rangle$ from the second to the first electrode.
Due to the charge conservation it can be calculated in any of the contacts,
say, $j=2$:
\begin{equation}\label{<I>_integral}
\langle I\rangle
= e\,\frac{2\gamma}{\delta}\int\limits_0^\mT \frac{dt}{\mT} \Bigl[p_1(t)\, w_-(U_2)-p_0(t)\, w_+(U_2)\Bigr].
\end{equation}

\section{Qualitative discussion}\label{Sec:Qualitative_Discussion}

To achieve the turnstile operation, the gate voltage is
driven periodically with frequency~$f$, which leads to a
time dependence of $\mu(t)$ via $n_g(t)$.
For simplicity we consider a symmetric square drive shape,
\begin{equation}\label{E_g}
\mu(t)=\left\{\begin{array}{ll}
-A_g,& 0<t<\mT/2,\\ A_g,& \mT/2<t<\mT,
\end{array}\right.
\end{equation}
where $\mT=1/f$ and $A_g$ are the drive period and amplitude, respectively.

The main idea of the electron turnstile \cite{SINIS_turnstile_Nat2008}
is to charge the island by transferring one electron from the first electrode
during the first half-period (injection stage), so that
$p_1(\mT/2)\to 1$, and to discharge it through
the second electrode in the second half-period,
$p_1(\mT)=p_1(0)\to 0$ (ejection stage).
Ideally, this should lead to a quantized value of the
period-averaged current through the device
$\langle I\rangle = ef$. 
Successful turnstile operation implies that the total charge
relaxation rate,
\begin{equation}\label{Gamma=}
\Gamma=\sum_j\frac{2\gamma}\delta
\left[w_+(U_j)+w_-(U_j)\right] \ ,
\end{equation}
well exceeds the drive frequency, $\Gamma\gg{f}$.
The smallness of $\delta\ll E_C,\Delta,eV$ which enabled us
to neglect $n$-dependence of $\mF_n(\ep)$  also leads to
separation between the relaxation time scales of $p_n$ and
$\mF(\ep)$.
Namely, the typical relaxation rate of the distribution function
$\mF(\ep)$ at a given energy due to tunneling is
$\Gamma_\mF\lesssim\gamma$ (an electron must
enter or leave a given level), while for $p_n$ the typical rate is
$\Gamma\sim\gamma(eV/\delta)\gg\Gamma_\mF$, as an
electron entering any of
$\sim eV/\delta\gg{1}$ levels is sufficient.
If, in addition, $\Gamma_\mF\ll{f}$, then one can solve
rate
equations~\eqref{rate_eq} assuming $\mF(\ep)$ to be fixed and
average Eq.~\eqref{kinetic_eq} over the relaxation time
of $p_n$ or over the drive period. In the opposite case one can
neglect the deviation of $p_n$ from the stationary solution (see Sec.~\ref{App:slow_drive}).

For the amplitude below the so-called forward-tunneling threshold
$A_g^{\rm ft}=\Delta-eV/2$ (we always assume $eV<2\Delta$)
the tunneling rates $w_\pm(U_j)$ are exponentially small.
Note that at finite bath temperatures $T$ one can still work in this regime at $A_g\simeq A_g^{\rm ft}-T$.
This leads to a refrigerating effect on the normal island which overheats superconducting electrodes.
In the following, we consider the amplitudes $A_g>A_g^{\rm ft}$,
to have a measurable turnstile current and to avoid overheating of the leads.
Still, several processes can lead to a deviation of
$\langle I\rangle$ from its ideal value $e f$.
Neglecting all high order processes in the tunneling rates (such as cotunneling, Andreev tunneling, and Cooper pair-electron tunneling)\cite{AverinPekola2008} and
focusing on the sequential tunneling contributions,
we can separate four types of errors related to
non-equilibrium effects both in the island and in
the superconducting leads:
\begin{multline}\label{<I>_general}
{\langle I\rangle} =
e f(1-2P_{\rm mt})(1-2P_\infty)\left(1-2 P_{\rm bt}\right)
+I_{\rm leak}.
\end{multline}
Here $P_{\rm mt}$ stands for the probability of missed
tunneling events if the inequality $\Gamma\gg{f}$ is
not strongly satisfied, so that at too large~$f$
electrons do not have enough time to tunnel into (out of)
the island during the corresponding half-period.
The factor $1-2P_\infty$ relates to the Pauli blocking
and differs from unity if a finite population of
quasiparticles (often non-equilibrium) is present in
the leads. The third factor $1-2 P_{\rm bt}$ in
Eq.~\eqref{<I>_general} is due to a combination of
back-tunneling processes and drive-dependent leakage
which produce tunneling events through the ``wrong''
junctions: through the first (second) one on the
ejection (injection) stage.
Finally, the leakage current $I_{\rm leak}$ is
present even
without drive ($A_g=0$) and is due to quasiparticles
which can tunnel through the device at any time and
increase the total current.
Typically, $I_{\rm leak}$ is independent of the
drive frequency.

The standard picture of the SINIS turnstile operation assumes
the electrons both on the island and in the leads to be in equilibrium with the
bath, whose temperature~$T$ is sufficiently low.
This ensures that the density of quasiparticle excitations is negligible, so that the quasiparticle leakage rate is small compared to the drive frequency.
Then all error contributions are small.
For this equilibrium assumption to be valid, the electron-phonon
relaxation time $\tau_{eph}$, must be short enough,
$\Gamma_{\mF}\tau_{eph}\ll 1$, where $\Gamma_\mF$
is the 
typical relaxation rate for the distribution
function $\mF(\ep)$ at a given energy due to the tunneling
terms in the kinetic equation~\eqref{kinetic_eq}.
In this equilibrium regime, the period-averaged current
$\langle I\rangle$ is quantized when the gate amplitude $A_g$
lies in the interval $A_g^{\rm ft}<A_g<A_g^{\rm bt}$
between the forward and backward tunneling thresholds,
$A_g^{\rm ft,bt}=\Delta\mp{e}V/2$ (see Fig.~\ref{Fig2:SINIS_turnstile}(b) for the energy diagram).
For $A_g>A_g^{\rm bt}$, the probability $P_{\rm bt}$ of tunneling
to the wrong junction becomes of order of unity.

\begin{figure}[h]
\center{
\includegraphics[width=0.4\textwidth]{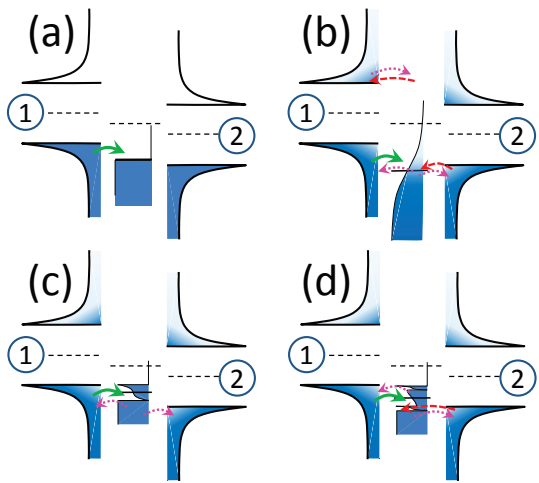}
}
\caption{(Color online)
The energy diagrams of a SINIS SET turnstile in the fast drive regimes mentioned in the text, when the drive frequency $f\gg \Gamma_{\mF}$,
the distribution function relaxation rate $\Gamma_{\mF}$ due to tunneling.
(a) Equilibrium regime $\Gamma_{\mF}\ll \tau_{eph}^{-1}$ at low bath temperature;
(b) Quasiequilibrium regime $\tau_{eph}^{-1}\ll\Gamma_{\mF}\ll \tau_{ee}^{-1}$ with the long thermal tails of the distribution function $\mF(\ep)$;
(c) and (d) Non-equilibrium regime $\Gamma_{\mF}\gg \tau_{ee}^{-1},\tau_{eph}^{-1}$ for (c) small $A_g<\Delta$ and (d) large $A_g>\Delta$ amplitudes.
In all panels the distribution functions in the island shown by blue color correspond to the first half-period ($\mu$ is shown by black solid lines).
The solid green, dashed red, and dotted purple arrows show the main forward tunneling, backtunneling, and leakage contributions, respectively.
}\label{Fig3:Energy_Diagrams}
\end{figure}
\begin{figure}[t]
\center{
\includegraphics[width=0.35\textwidth]{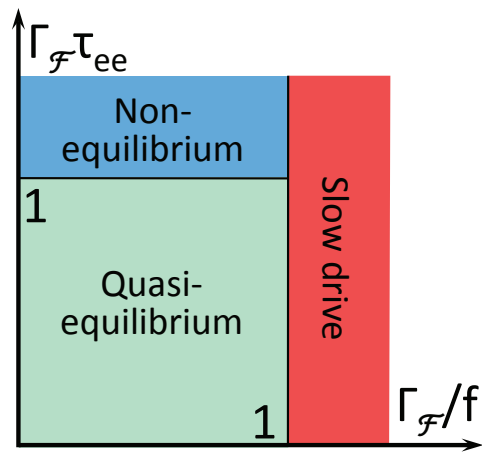}
}
\caption{(Color online)
A schematic diagram of different regimes
of a SINIS SET turnstile out of equilibrium, $\tau_{eph}^{-1}\ll\Gamma_{\mF}$, showing the three regimes discussed in the text.
}\label{Fig4:Diagram}
\end{figure}
\begin{figure}[t]
\center{
\includegraphics[width=0.4\textwidth]{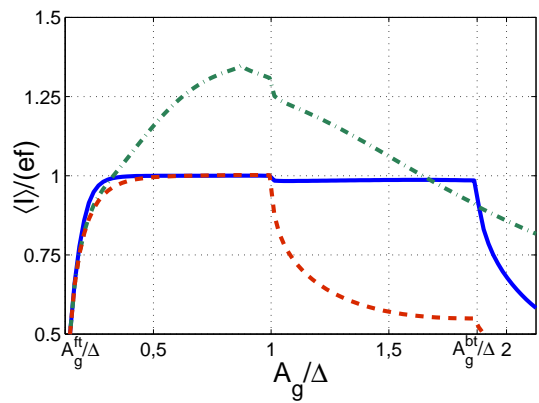}
}
\caption{(Color online)
The period-averaged current $\langle I \rangle$ through the SINIS SET versus the drive amplitude $A_g/\Delta$
for the equilibrium (blue solid line), quasiequilibrium (green dashed-dotted line) and fully non-equilibrium (red dashed line) regimes
discussed in the text, and $e V = 1.75\,\Delta$.
The threshold values, $A_g^{\rm ft} = 0.125\,\Delta$ and $A_g^{\rm bt}=1.875\,\Delta$, are shown explicitly on the horizontal axis.
The calculation method and parameter values are discussed below in Sec.~\ref{Sec:Estimates} and in Fig.~\ref{Fig6:I_ef}.
The finite widths of the current plateau onsets at small $(A_g-A_g^{\rm ft})/\Delta\lesssim 0.2$ are related to the missed tunneling events
due to the regime $\Gamma/f\lesssim 1$.
}\label{Fig5:I_ef_3limits}
\end{figure}

In this work, we investigate the turnstile operation at
$\Gamma_{\mF}\tau_{eph}\gtrsim 1$, i.~e., beyond the
equilibrium limit, and demonstrate that even in this case
there is a certain range of parameters where the SINIS SET can
still work as an electronic turnstile.
Specifically, beyond the equilibrium limit we identify three
regimes with different behavior of the electron distribution
function $\mF(\ep,t)$ in the normal island.
\renewcommand{\labelenumi}{(\roman{enumi})}
\begin{enumerate}
\item
In the quasiequilibrium regime (see the energy diagram Fig.~\ref{Fig3:Energy_Diagrams}(b) and the light green area close to the origin in Fig.~\ref{Fig4:Diagram}),
$\Gamma_\mF\ll{f},1/\tau_{ee}$, the distribution function
$\mF(\ep,t)\approx \mF_{T_e}(\ep)$ has the Fermi-Dirac
form with an effective electronic temperature
$T_e\lesssim A_g-A_g^{\rm ft}$, governed by the balance of the
heat flows. In this regime the current deviates significantly from
$e f$ (except a narrow range, $A_g/A_g^{\rm ft}-1\ll 1$, of the parameters near the threshold) due to long thermal tails of $\mF_{T_e}(\ep)$
as shown by green dash-dotted line in Fig.~\ref{Fig5:I_ef_3limits}.
\item
In the fully non-equilibrium regime (see the energy diagrams Fig.~\ref{Fig3:Energy_Diagrams}(c, d) and the blue area on the top of Fig.~\ref{Fig4:Diagram}),
$1/\tau_{ee}\ll\Gamma_\mF\ll{f}$,
the tunneling processes form a non-Fermi-Dirac energy profile
of $\mF(\ep)$. This regime is characterized by a wide
distribution $\mF(\ep)$ with sharp jumps at
$\ep=\pm (A_g-A_g^{\rm ft})$. The sharpness of these jumps,
determined by the smearing of the Bardeen-Cooper-Schrieffer (BCS) singularities, 
ensures good current quantization $\langle I\rangle\simeq e f$ (see Fig.~\ref{Fig5:I_ef_3limits})
in the gate amplitude interval $A_g^{\rm ft}<A_g<\Delta$
(see Fig.~\ref{Fig3:Energy_Diagrams}(c)), despite a strongly
non-equilibrium shape of the distribution function.
\item
At slow drive (see the dark red area on the right in
Fig.~\ref{Fig4:Diagram} and the energy diagram in Fig.~\ref{Fig9:Slow_drive}), $f\ll \Gamma_{\mF}$,
the distribution function may evolve significantly over the drive
period $\mT=1/f$.
This regime is characterized by the large leakage current due
to the relaxation of $\mF(\ep,t)$ by electron tunneling
processes.
\end{enumerate}

In the next section we focus on the case of the fast drive and study cases (i) and (ii).
We will also consider the equilibrium regime as a reference point.
The slow drive limit is studied separately in Sec.~\ref{App:slow_drive}.
%

\section{Fast drive}
\label{Sec:Fast_drive}

Here we take advantage of the assumption $\Gamma_\mF\ll{f}$,
covering cases (i) and (ii), discussed in the previous section.
This assumption enables us to neglect the change in $\mF(\ep)$
during the period in the first approximation, and write down
the solution for $p_1(t)=1-p_0(t)$ as
\begin{subequations}\label{p_n_sol}
\begin{align}
\label{p_n_sol_1st_half}p_1(t<\mT/2) &= 1-P_\infty - \left(1-2P_\infty\right)\frac{e^{-\Gamma t}}{1+e^{-\Gamma\mT/2}} \ , \\
p_1(t>\mT/2) &= P_\infty+\left(1-2P_\infty\right)\frac{e^{-\Gamma(t-\mT/2)}}{1+e^{-\Gamma\mT/2}} \ ,
\end{align}
\end{subequations}
where we consider $0<t<\mT$.
The quantity $P_\infty$ is specified below.
Both $\Gamma$ and $P_\infty$ depend on $\mF(\ep)$. For the moment,
let us consider $\mF(\ep)$ as given, it will be
determined later from the kinetic equation.

The total charge relaxation rate $\Gamma$ in Eqs.~(\ref{p_n_sol}),
defined by Eq.~(\ref{Gamma=}), is a sum of four terms,
$\Gamma=\Gamma_{1+}+\Gamma_{1-}+\Gamma_{2+}+\Gamma_{2-}$, with $\Gamma_{j\pm} = 2\gamma w_\pm(U_j)/\delta$.
As long as $\mF(\ep)$ is assumed to be constant during the
period, $\Gamma$~is also constant.
This is the consequence of the electron-hole symmetry of the
problem, $\mF(\ep)=1-\mF(-\ep)$,
following from the symmetry of the drive~(\ref{E_g}).
Indeed, each $\Gamma_{j\pm}$ is constant during each
half-period, but different on the two half-periods because
of the dependence $\mu(t)$, Eq.~(\ref{E_g}). The symmetry
of the latter, however, leads to the relations
\begin{subequations}\label{Gamma12pm=}
\begin{align}
&\Gamma_{1\pm}(0<t<\mT/2)=
\Gamma_{2\mp}(\mT/2<t<\mT),\\
&\Gamma_{2\pm}(0<t<\mT/2)=
\Gamma_{1\mp}(\mT/2<t<\mT),
\end{align}\end{subequations}
so the sum of the four terms is the same on both half-periods.
In the following, we will omit the time arguments, referring
to the values on the first half-period.

$P_\infty$ in Eqs.~(\ref{p_n_sol}) represents the limiting
probability of having $n=0$ or~1 excess electrons on the
island at the end of the injection or ejection stage,
respectively, at long enough times, $\Gamma \mT\gg 1$.
It is given by
\begin{subequations}\label{errors=}
\begin{equation}\label{P_infty}
P_\infty =\frac{\Gamma_{1-}+\Gamma_{2-}}\Gamma.
\end{equation}
Out of four terms $\Gamma_{j\pm}$, only $\Gamma_{1+}$ is large
in the equilibrium turnstile operation regime,
$A_g^{\rm ft}<A_g<A_g^{\rm bt}$, while the rest are due
to residual quasiparticles in the leads and thermal tails of the distribution function in the island, so $P_\infty\ll{1}$.
Substituting solution~(\ref{p_n_sol}) into expression
(\ref{<I>_integral}), we obtain \eqref{<I>_general} with
$P_\infty$ written above and other error contributions given by
\begin{align}
\label{P_mt}
&P_{\rm mt} =
\frac{e^{-\Gamma\mT/2}}{1+e^{-\Gamma\mT/2}},\\
&P_{\rm bt} = \frac{\Gamma_{2+}+\Gamma_{2-}}\Gamma,\\
&I_{\rm leak} = e\,\frac{\Gamma_{1+}\Gamma_{2-}-\Gamma_{1-}\Gamma_{2+}}\Gamma.
\label{I_leak}
\end{align}
\end{subequations}

\begin{figure}[h]
\center{
\includegraphics[width=0.4\textwidth]{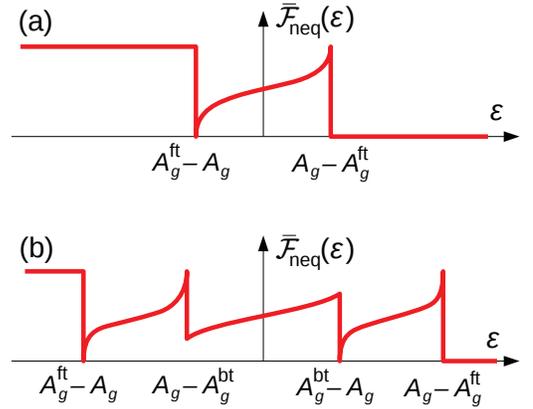}
}
\caption{(Color online)
A sketch of the dependence $\bar\mF_{\rm neq}(\ep)$ in the
turnstile regime $A_g^{\rm ft}<A_g<A_g^{\rm bt}$ for
(a)~$A_g<\Delta$ and (b)~$A_g>\Delta$.
}
 \label{Fig5:Fneq}
\end{figure}

To determine the distribution function $\mF(\ep)$
which implicitly enters the expressions above, we average
Eq.~(\ref{kinetic_eq}) over the period with the help of
Eqs.~(\ref{E_g}) and~(\ref{p_n_sol}) (as discussed above,
the evolution of $\mF(\ep)$ is slow). Collecting
the terms linear in $\mF(\ep)$ and independent of $\mF(\ep)$
in the tunneling part of Eq.~(\ref{kinetic_eq}), we can rewrite
it identically as
\begin{gather}\label{kinetic_eq_aver_f_nu}
\dot{\mF}(\ep) = \bar\Gamma_{\mF}(\ep)
\left[\bar\mF_{\rm neq}(\ep)-\mF(\ep)\right]+\St[\mF](\ep),
\end{gather}
where
$\bar\Gamma_{\mF}(\ep)=\gamma [\bar w_{\mF}(\ep)+\bar w_{\mF}(-\ep)]$
is the period-averaged energy-dependent rate of
relaxation towards a certain non-equilibrium distribution
$\bar\mF_{\rm neq}(\ep)=\gamma \bar w_{\mF}(\ep)/\bar\Gamma_{\mF}(\ep)$,
determined by the tunneling to and from the electrodes.
Here we defined
\begin{multline}
\bar w_{\mF}(\ep) = \sum_{\eta=\pm}\left[\left(\frac{1}{2}-\bar p\right) n_S^+(A_g+\eta eV/2-\ep) +\right.\\ 
\left.\phantom{\frac{1}{2}}\bar p\,n_S^+(-A_g+\eta eV/2-\ep)\right] 
 \ ,
\end{multline}
\begin{gather}\label{n_S^+}
n_S^+(\ep) = n_S\left(\ep\right)\left[1-\mF_T\left(\ep\right)\right]\approx n_S\left(\ep\right)\theta\left(\ep\right) \ ,
\end{gather}
\begin{multline}\label{p_n_aver}
\bar{p} = \frac{1}{\mT}\int_0^{\mT/2} p_1(t) dt = \frac{1}{\mT}\int_{\mT/2}^\mT p_0(t) dt = \\
{}=\frac{1-P_\infty}{2}-\frac{(1-2P_\infty)(1-2P_{\rm mt})}{\Gamma \mT} \ .
\end{multline}
The distribution $\bar\mF_{\rm neq}(\ep)$ is very far from
the equilibrium one. Although it tends to 0 and 1 in the limits
$\ep\to\pm\infty$, in the interval $|\ep|<A_g-A_g^{\rm ft}$
it has several sharp features, inherited from the BCS singularities
in the electrode density of states.
$\bar\mF_{\rm neq}(\ep)$ may have different shape, depending
on how these singularities, shifted by $\pm{A}_g$ and/or
$\pm{e}V/2$ are located with respect to each other.
In the turnstile regime, $A_g^{\rm ft}<A_g<A_g^{\rm bt}$,
it may have two possible shapes which are shown schematically in
Fig.~\ref{Fig5:Fneq}.

The relaxation rate
$\bar\Gamma_{\mF}(\ep)$ competes with the collision integral
which tends to drive the system towards equilibrium.
When $\bar{p}$ is close to~$1/2$, $\bar\Gamma_{\mF}(\ep)$
has quite different values, depending on whether the second term
in the expression $\bar w_{\mF}(\ep)$ vanishes or not. Namely,
$\bar\Gamma_{\mF}(\ep)\sim\gamma$ for energies
$|\ep|>A_g+A_g^{\rm ft}$, which corresponds to
ejecting electrons from the island into the empty QP bands on
the injection stage, $0<t<\mT/2$, or to electron
injection from the deep filled QP states on the ejection stage,
$\mT/2<t<\mT$. These processes contribute to
the errors via the rates $\Gamma_{1-},\Gamma_{2-}$, which are
suppressed when $\mF(\ep)$ is close to 0 or 1 at
$|\ep|>A_g+A_g^{\rm ft}$.

In the energy interval $|\ep|<A_g+A_g^{\rm ft}$, more
important for the turnstile operation, the relaxation rate
is smaller, $\bar\Gamma_\mF(\ep)\sim\Gamma_\mF$,
which we define as
\begin{equation}\label{Gamma_f}
\Gamma_\mF\equiv\gamma\left(\frac{1}{2}-\bar p\right)\approx
\frac{\gamma{P}_\infty}2+\frac{f\gamma}{\Gamma}.
\end{equation}
Here the first contribution $\gamma P_\infty$ is related to
the time-independent terms in Eq.~\eqref{p_n_sol}, while
$f \gamma/\Gamma$ is the effective rate associated with fast
charge transfer terms in Eq.~\eqref{p_n_sol} and originated from
the second term in Eq.~\eqref{p_n_aver}.
Because the interval $|\ep|<A_g+A_g^{\rm ft}$ is the most important for the turnstile operation,
it is $\Gamma_\mF$ from Eq.~\eqref{Gamma_f} that should be used in the validity condition of Eq.~\eqref{kinetic_eq_aver_f_nu}.
In the opposite limit of $\max(\gamma P_\infty,1/\tau_{ee})\gg f$, considered in Sec.~\ref{App:slow_drive}, one can neglect the charge relaxation in $p_n(t)$ in the kinetic equation, i.e., in the time-dependent terms in Eq.~\eqref{p_n_sol}.

In the $\tau$-approximation for the collision integral,
Eqs.~\eqref{St=},
the steady-state solution of Eq.~(\ref{kinetic_eq_aver_f_nu})
can be written explicitly:
\begin{gather}\label{f_stat_sol}
\mF(\ep) = \frac{\bar\Gamma_{\mF}(\ep)\, \bar\mF_{\rm neq}(\ep)+\mF_{T_e}(\ep)/\tau_{ee}+\mF_{T}(\ep)/\tau_{eph}}{\bar\Gamma_{\mF}(\ep)+1/\tau_{ee}+1/\tau_{eph}} \ .
\end{gather}
Note that due to the symmetry of the drive we have $\bar\Gamma_{\mF}(\ep)=\bar\Gamma_{\mF}(-\ep)$ and $\bar\mF_{\rm neq}(\ep)=1-\bar\mF_{\rm neq}(-\ep)$ leading to $\mF(\ep)=1-\mF(-\ep)$ and $w_-[U]=w_+[-U]$.

Below we consider the limiting cases corresponding to the equilibrium ($\Gamma_{\mF}\ll\tau_{eph}^{-1}$), quasiequilibrium ($\tau_{eph}^{-1}\ll\Gamma_{\mF}\ll\tau_{ee}^{-1}$) and fully non-equilibrium ($\Gamma_{\mF}\gg\tau_{ee}^{-1}$) regimes.
Here the typical value of the distribution function relaxation rate $\Gamma_{\mF}$ \eqref{Gamma_f} corresponds to the energy interval $|\ep|<A_g-A_g^{\rm ft}$ where the major deviations of $\mF(\ep)$ from the equilibrium one occur.
In the following, we show that the plateau $\langle I\rangle = e f$ of the pumping current survives in the equilibrium and, surprisingly, in the fully non-equilibrium regimes, but it is destroyed by the thermal tails of the distribution function in the quasiequilibrium one.
Such unusual non-monotonic behavior demonstrates the negative role of the electron-electron relaxation for the turnstile operation, but not of non-equilibrium itself.

\subsection{Equilibrium regime}\label{Sec:Eq}

For reference, we consider first the standard equilibrium regime, when
the fast enough electron-phonon relaxation with the rate
$\tau_{eph}^{-1}\gg\Gamma_{\mF}$ drives the system towards equilibrium
with the phonon bath (whatever the relation between $\tau_{ee}$ and
$\tau_{eph}$).
In this case, one can neglect the difference between $\mF(\ep,t)$
and the equilibrium Fermi distribution $\mF_T(\ep)$ with the small
bath temperature $T\ll \Delta, E_C$.
In the turnstile operation regime,
$A_g^{\rm ft}<A_g<A_g^{\rm bt}$, only $\Gamma_{1+}$ is not
exponentially small:
\begin{subequations}\begin{eqnarray}
&&\Gamma_{1+}=\frac{2\gamma}\delta\,\sqrt{(A_g+eV/2)^2-\Delta^2},\\
&&\Gamma_{2+}=\frac{2\gamma}\delta\,\sqrt{\frac{\pi{T}\Delta}{2}}\,
e^{-(A_g^{\rm bt}-A_g)/T},
\end{eqnarray}\end{subequations}
while $\Gamma_{1-}$ and $\Gamma_{2-}$ are suppressed by even stronger
exponentials $e^{-(A_g+eV/2)/T}$ and $e^{-\max\{\Delta,A_g-e V/2\}/T}$, respectively, not exceeding $e^{-\Delta/T}$ at $A_g>A_g^{\rm ft}$.
As a result,
both the leakage and the Pauli blocking errors are negligible,
and the errors are dominated by the backtunneling,
\begin{equation}
P_{\rm bt}\approx\frac{\Gamma_{2+}}{\Gamma_{1+}}
=\sqrt{\frac{(\pi/2){T}\Delta}{(A_g+eV/2)^2-\Delta^2}}\,
e^{-(A_g^{\rm bt}-A_g)/T}
\end{equation}
(we assume the missed tunneling contribution to be well suppressed
by the exponential $e^{-\Gamma\mT}$).
As soon as the drive amplitude exceeds the standard back-tunneling
threshold $A_g^{\rm bt}=\Delta+eV/2$, the current $\langle I\rangle$
starts to strongly deviate from the plateau $e f$.

One may consider corrections to this result due to the deviation
of $\mF(\ep)$ from $\mF_T(\ep)$, small by the factors
$\Gamma_\mF\tau_{eph}$, $\tau_{eph}/\tau_{ee}$. Contributions
to the errors from the term with $\mF_{T_e}(\ep)$ are exponentially
small because the electronic temperature $T_e$, found
self-consistently, is close to $T$ (by the same small factors).
Contributions originating from $\bar\mF_{\rm neq}(\ep)$ can be calculated analogously
to the ones in Sec.~\ref{Sec:Slow_ee_relax};
here we note that for $A_g<\Delta$ they are also exponentially
small as $e^{-\Delta/T}$, while for $A_g>\Delta$ the smallness
is no longer exponential and is guaranteed only by the parameter
$\Gamma_\mF\tau_{eph}$.

\subsection{Quasiequilibrum regime}\label{Sec:Fast_ee_relax}

When $\tau_{eph}^{-1}\ll\Gamma_{\mF}\ll \tau_{ee}^{-1}$,
one can separate $\mF_{T_e}(\ep)$ as the main contribution
to the distribution function \eqref{f_stat_sol},
\begin{align}
\mF(\ep) \approx \mF_{T_e}(\ep)
&+\tau_{ee}\bar\Gamma_{\mF}(\ep)
\left[\bar\mF_{\rm neq}(\ep)-\mF_{T_e}(\ep)\right]+{}\nonumber\\
&{}+\frac{\tau_{ee}}{\tau_{eph}}\left[\mF_{T}(\ep)-\mF_{T_e}(\ep)\right],
\label{f_stat_sol_fast_tee}
\end{align}
and naturally rewrite Eq.~\eqref{T_e} for the electronic
temperature $T_e$ as the heat balance equation:
\begin{align}
&\frac{2}\delta\int \ep\,\bar\Gamma_{\mF}(\ep)
\left[\bar\mF_{\rm neq}(\ep)-\mF_{T_e}(\ep)\right]d\epsilon
=\nonumber\\
&=\frac{2}\delta
\int \ep\,\frac{\mF_{T_e}(\ep)-\mF_{T}(\ep)}{\tau_{eph}}\,d\ep.
\label{heat_balance}
\end{align}
In fact, the right-hand side of this equation is nothing but the
total cooling power~(\ref{W=St}), and in the limit
$\Gamma_{\mF}\ll \tau_{ee}^{-1}$ this right-hand side can be replaced by the more
precise dependence (\ref{coolingpower=}).
Moreover, one can actually take the limit $\tau_{eph}\to\infty$
and set the right-hand side of Eq.~(\ref{heat_balance}) to zero.
Even in this case, a finite value of $T_e$ is obtained, which
satisfies the following equation:
\begin{equation}
\int\ep\,\bar w(-\ep)\,\mF_{T_e}(\ep)\,d\ep=0.
\end{equation}
This equation has a unique solution $T_e\sim{A}_g-A_g^{\rm ft}$.

Near the forward tunneling threshold,
$0<A_g-A_g^{\rm ft}\ll{e}V$,
the temperature $T_e=(A_g-A_g^{\rm ft})/c_0$,
with $c_0=0.7233\ldots$, and we can evaluate
\begin{subequations}\begin{align}
\Gamma_{1+}={}&{}\frac{2\gamma}\delta
\int\limits_{A_g^{\rm ft}-A_g}^\infty
n_S^+(\ep+A_g+eV/2)\,\mF_{T_e}(\ep)\,d\ep\approx\nonumber\\
\approx{}&{}c_1\,\frac{2\gamma}\delta\sqrt{(A_g-A_g^{\rm ft})\Delta},\\
\Gamma_{2+}={}&{}\frac{2\gamma}\delta
\int\limits_{A_g^{\rm bt}-A_g}^\infty
n_S^+(\ep+A_g-eV/2)\,\mF_{T_e}(\ep)\,d\ep\approx\nonumber\\
\approx{}&{}c_2\,\frac{2\gamma}\delta\sqrt{(A_g-A_g^{\rm ft})\Delta}\,
e^{-c_0eV/(A_g-A_g^{\rm ft})},
\end{align}\end{subequations}
where the numerical factors are approximately given by
$c_1=1.3329\ldots$, $c_2=6.0751\ldots$.
Other rates are suppressed by even stronger exponentials,
$\Gamma_{1-,2-}\sim\Gamma_{1+}
e^{-c_0(A_g+\Delta\pm{e}V/2)/(A_g-A_g^{\rm ft})}$.\cite{Footnote_QEq_small_rates}
However, away from the threshold, when $A_g-A_g^{\rm ft}\sim{e}V$,
the electronic temperature rapidly increases and so do
all errors. Thus, in most of the quasiequilibrium regime,
the electron-electron collisions are detrimental for the
turnstile operation.

\subsection{Fully non-equilibrium regime}\label{Sec:Slow_ee_relax}

When $\tau_{eph}^{-1},\tau_{ee}^{-1}\ll\Gamma_{\mF}$,
the distribution function can be written as
\begin{gather}\label{f_stat_sol_slow_tee}
\mF(\ep)=\bar\mF_{\rm neq}(\ep)+
\frac{\St[\bar\mF_{\rm neq}](\ep)}{\bar\Gamma_{\mF}(\ep)}
+O\left(\frac{\St[\bar\mF_{\rm neq}]^2}{\bar\Gamma_{\mF}^2}\right).
\end{gather}
Let us start by analyzing the main term, $\bar\mF_{\rm neq}(\ep)$,
and its contribution to the turnstile errors. In the range
$A_g^{\rm ft}<A_g<\Delta$, the distribution
$\bar\mF_{\rm neq}(\ep)$ has the shape, shown schematically in
Fig.~\ref{Fig5:Fneq}(a) and is given by the simple expression,
\begin{gather}\label{f_nu_simple}
\bar\mF_{\rm neq}(\ep) =\frac{n_S^+(A_g+\tfrac{eV}2- \ep)}{n_S^+(A_g+\tfrac{eV}2- \ep)+n_S^+(A_g+\tfrac{eV}2+\ep)}.
\end{gather}
A similar behavior of the distribution function was predicted theoretically for the dc mode of an analogous setup beyond the subgap regime $eV>2\Delta$
in Ref.~\onlinecite{Pekola_Heikkila_Non-eq_f(E)-2},
as well as for an entirely superconducting device in Ref.~\onlinecite{Heikkila_SISIS_non-eq_f(E)}
This determines the dominant among the four rates $\Gamma_{j\pm}$,
\begin{align}
&\Gamma_{1+}=\frac{2\gamma}{\delta}\int\limits_\Delta^{2A_g-\Delta+eV}
\frac{n_S(\ep)\,n_S(2A_g+eV-\ep)}{n_S(\ep)+n_S(2A_g+eV-\ep)}\,d\ep=\nonumber\\
&\qquad{}=\frac{2\gamma}\delta\,\sqrt{2(A_g-A_g^{\rm ft})\Delta}
\left[\sqrt{2}+\ln(\sqrt{2}-1)\right]+{}\nonumber\\
&\qquad\quad{}+\frac{2\gamma}\delta\,
O\!\left((A_g-A_g^{\rm ft})^{3/2}/\sqrt{\Delta}\right),
\label{Gamma1+neq}
\end{align}
where we only give an explicit expression near the threshold,
$A_g-A_g^{\rm ft}\ll\Delta$ (although the integral can be
calculated exactly, the resulting expressions are quite bulky
and not very informative). The rest of the rates,
$\Gamma_{2+},\Gamma_{1-},\Gamma_{2-}$, are all exponentially
suppressed as $\propto{e}^{-\Delta/T}$, where $T$ is the bath
temperature, as they are determind by the quasiparticle
population in the superconducting electrodes. This happens
because at $A_g^{\rm ft}<A_g<\Delta$, the energy range
where $\bar\mF_{\rm neq}(\ep-U_j)$ in
Eqs.~(\ref{w+=}),~(\ref{w-=}) for $\Gamma_{2+},\Gamma_{1-},\Gamma_{2-}$ is different from 0 or 1,
falls inside the superconducting gap in $n_S(\ep)$.

The situation changes dramatically at $A_g>\Delta$, when the
energy interval of $\bar\mF_{\rm neq}(\ep)$ widens, and a
second peak-dip structure appears, as shown in Fig.~\ref{Fig5:Fneq}(b).
Then, backtunneling becomes allowed and its rate $\Gamma_{2+}$
is no longer exponentially small; near the threshold,
$0<A_g-\Delta\ll\Delta,eV$, the rate $\Gamma_{2+}$ is given by
the same expression~(\ref{Gamma1+neq}) with the replaced threshold,
$A_g^{\rm ft}\to\Delta$. Thus, at $A_g>\Delta$, the turnstile
accuracy quickly drops.

We now turn back to the ``good'' region, $A_g^{\rm ft}<A_g<\Delta$,
and study errors originating from the subleading term in
Eq.~(\ref{f_stat_sol_slow_tee}). These errors are important as they
have no exponential smallness $e^{-\Delta/T}$, in contrast to those
originating from the main term.
Indeed, the temperature $T_e$ determining the electron-electron
part of the collision integral~(\ref{Stee=}), is not necessarily
low. Up to small corrections, $T_e$~can be found from the equation
\begin{equation}
\int \ep\left[\bar\mF_{\rm neq}(\ep)-\theta(-\ep)\right]d\ep
=\frac{\pi^2T_e^2}6.
\end{equation}
Quite analogously to the quasiequilibrium case, this gives
$T_e\sim{A}_g-A_g^{\rm ft}$.
We can give an analytical expression for the main error
only near the forward threshold, $0<A_g-A_g^{\rm ft}\ll{e}V$:
\begin{align}
&\Gamma_{2+}\approx\frac{c_2'}{(1/2-\bar{p})\tau_{ee}}\,
\frac{A_g-A_g^{\rm ft}}{\delta}\,e^{-c_0'eV/(A_g-A_g^{\rm ft})},\\
&P_{\rm bt}\approx\frac{c_2'\mT}{\tau_{ee}}\,
\frac{A_g-A_g^{\rm ft}}{\delta}\,e^{-c_0'eV/(A_g-A_g^{\rm ft})},
\end{align}
with the numerical factors $c_0'=1.52\ldots$,
$c_2'=6.01\ldots$. Here we assumed that the main
contribution to $1/2-\bar{p}$ comes from
$1/(\Gamma\mT)$, rather than from~$P_\infty$
[see Eq.~(\ref{p_n_aver})].

\subsection{Parameter estimates and crossovers}\label{Sec:Estimates}

\begin{figure*}[t]
\center{
\includegraphics[width=0.95\textwidth]{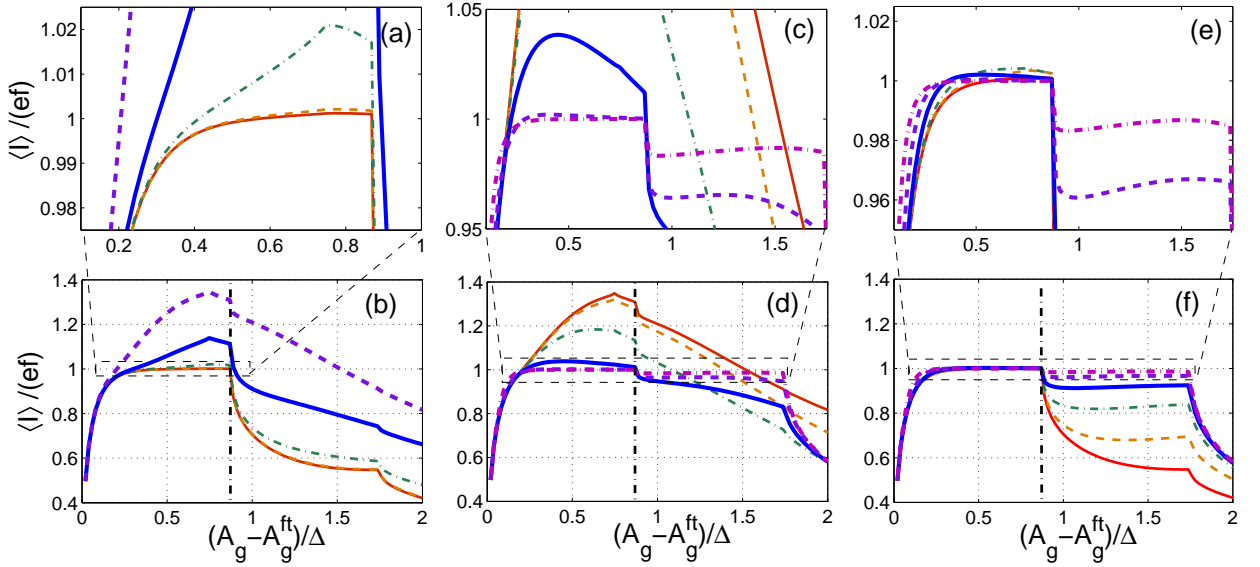}
}
\caption{(Color online) The period-averaged current $\langle I \rangle$ through the SINIS SET versus the drive amplitude $(A_g-A_g^{\rm ft})/\Delta$
relative to the threshold value for the equilibrium (Eq), $\Gamma_\mF\ll\tau_{eph}^{-1}$, quasiequilibrium (QEq),
$\tau_{eph}^{-1}\ll \Gamma_\mF\ll \tau_{ee}^{-1}$, and fully non-equilibrium (NEq), $\tau_{eph}^{-1}, \tau_{ee}^{-1}\ll \Gamma_\mF$, regimes
and crossovers between them.
In panels (a), (c), and (e) we represent zooms of the regions shown by dashed rectangles in panels (b), (d), and (f)
near the plateau $\langle I \rangle = e f$.
(a), (b)~NEq~--~QEq crossover 
governed by the decrease of $l_{\rm mfp}$ from $1$ (red thin solid) to $2.4\cdot 10^{-2}$ (violet thick dashed-dotted) by a factor $\sqrt{10}$ at each step.
This corresponds to the change of the Thouless energy $E_{\rm Th}/\Delta$ from $75$ (red thin solid) to $1.8$ (violet thick dashed-dotted).
We take $\delta/\Delta = 10^{-2}$ keeping $\Gamma_\mF$ constant.
(c), (d)~QEq~--~Eq crossover 
governed by the decrease of $\delta/\Delta$ from $10^{-2}$ (red thin solid) to $10^{-7}$ (violet thick dashed-dotted curve) by a factor $10$ at each step.
We choose $l_{\rm mfp}\propto (\delta/\Delta)^{-1/3}$ and change it from $2.4\cdot 10^{-2}$ up to $1$
to keep the Thouless energy constant, $E_{\rm Th}/\Delta =1.8$.
This effectively changes only $\Gamma_\mF\simeq f\delta/\Delta$.
In all curves $\tau_{ee}^{-1}\gg \Gamma_\mF$.
(e), (f)~NEq~--~Eq crossover 
governed by the decrease of $\delta/\Delta$ from $10^{-2}$ (red thin solid) to $10^{-7}$ (violet thick dashed-dotted curve) by a factor $10$ at each step.
We consider the maximal possible Thouless energy taken at $l_{\rm mfp}=1$, however, due to $\delta$-dependence of $E_{\rm Th}/\Delta$,
it also varies from $75$ (red thin solid) to $1.8$ (violet thick dashed-dotted).
In all panels we considered the island made of AlMn ($T_{eph}=250$~K, $E_F=11.7$~eV and $\nu_\mathcal{N}(0)=1.45\cdot 10^{47}$~J$^{-1}$~m$^{-3}$) and
take $e V=1.75\Delta$, $f/\Delta=2\cdot 10^{-3} \simeq 300$~MHz, $\gamma/\delta=4\cdot 10^{-2}\gg f/\Delta$.
In panels (b), (d), and (f) the vertical dashed-dotted line shows the amplitude threshold value $A_g = \Delta$ for the non-equilibrium case.
The finite widths of the current plateau onsets at small
$(A_g-A_g^{\rm ft})/\Delta\lesssim 0.2$ are related to the missed tunneling events due to the regime $\Gamma/f\lesssim 1$.
}
 \label{Fig6:I_ef}
\end{figure*}

To illustrate the three regimes, described above, we present in Fig.~\ref{Fig6:I_ef}
the dependence of the turnstile current $\langle I\rangle/(ef)$
on the drive amplitude, $(A_g-A_g^{\rm ft})/\Delta$, relative to the threshold value,
which is obtained by the direct numerical evaluation of Eq.~(\ref{<I>_general})
via Eqs.~(\ref{errors=}) with the rates $\Gamma_{j\pm}$ evaluated
using the distribution function~(\ref{f_stat_sol}).

First, we estimate the possible parameters for each of the cases.
Typical frequencies of the turnstile operation are usually limited from above by the missed tunneling events \eqref{P_mt}
and the overheating of superconducting leads as $f\lesssim300$~MHz.
Here we consider electrodes made of aluminum and take into account some reduction of the superconducting gap due to the possible usage of quasiparticle traps and/or active cooling taking $\Delta= 1$~K.
Then, considering $f=300$~MHz, we can still neglect the finite QP density in the leads due to a exponentially small prefactor
$e^{-\Delta/T}$ and take into account only the second term in \eqref{Gamma_f} giving $\Gamma_\mF\sim f \delta/\Delta$.
Here and further we assume $A_g,e V\sim \Delta$ and fix the ratio $\gamma/\delta$ to the value $4\cdot 10^{-2}$ ensuring $\Gamma\sim \gamma\Delta/\delta\gg f\simeq 2\cdot 10^{-3}\Delta$.

The parameter $T_{eph}$ 
governing electron-phonon relaxation rate \eqref{taueph=}
depends only on the material of the normal island
and varies between $\sim60$~K for Au and $\sim 250$~K for AlMn.\cite{Giazotto2006,Footnote_AlMn}
The Thouless energy governing electron-electron relaxation \eqref{SIA=}
$E_{\rm Th}\sim\min\{v_F/L,D/L^2\}\sim E_F^{2/3}\delta^{1/3}l_{\rm mfp}$
can be expressed in terms of the mean level spacing $\delta$ determined by the island size $L\sim \mathcal{V}^{1/3}$,
of the Fermi energy $E_F\sim 1-10$~eV (for typical metals) through the Fermi velocity $v_F$, and of the mean free path $\ell$ in the island normalized to its size $l_{\rm mfp} = \min\{1,\ell/L\}$.

To achieve the fully non-equilibrium case (ii) one needs both following ratios to be large:
\begin{subequations}\label{Gteph+Gtee}
\begin{align}
\Gamma_\mF \tau_{eph} &= \frac{f\delta}{\Delta}\frac{ T_{eph}^2}{T_e^3}\gg 1 \ ,\\
\Gamma_\mF \tau_{ee} &= \frac{f}{\Delta}\frac{E_{\rm Th}^2}{T_e^2}\sim\frac{f}{\Delta}\frac{E_F^{4/3}\delta^{2/3}l_{\rm mfp}^2}{T_e^2} \gg 1 \ ,
\end{align}
\end{subequations}
with $T_e\sim A_g-A_g^{\rm ft}\sim \Delta$.
To maximize $T_{eph}=250$~K and $l_{\rm mfp}=1$ we assume the island to be made of AlMn
with a sufficiently low Mn concentration ensuring $\ell>L$, $E_F\simeq 11.7$~eV, and $\nu_\mathcal{N}(0)=1.45\cdot 10^{47}$~J$^{-1}$~m$^{-3}$.
After this material optimization we are left with the only free parameter $\delta$ to make the ratios large.
Within the assumption of continuous spectrum $\delta/\Delta\ll 1$ we consider $\delta = 10^{-2}\Delta$
corresponding to the island size $L\sim \mathcal{V}^{1/3}\simeq 50$~nm achievable with existing experimental techniques.\cite{SINIS_turnstile_high_EC}
This gives $\Gamma_\mF \tau_{eph}\sim 100$ and $\Gamma_\mF \tau_{ee}\sim 200$ for the typical electronic temperature in this case $T_e\simeq 0.25 \Delta$.
The corresponding plot of $\langle{I}\rangle/(ef)$ vs $(A_g-A_g^{\rm ft})/\Delta$ is shown
by the red dashed curve in Fig.~\ref{Fig5:I_ef_3limits} and
by the red thin solid curves in panels (a), (b), (e), (f) of Fig.~\ref{Fig6:I_ef}.

By reducing the mean free path $l_{\rm mfp}$ in the island from unity, one enhances the electron-electron relaxation rate
keeping the electron-phonon rate intact, so that the system crosses from the fully non-equilibrium case
(ii), $\tau_{ee}^{-1},\tau_{eph}^{-1}\ll \Gamma_\mF$, to the quasi-equilibrium case (i), $\tau_{eph}^{-1}\ll \Gamma_\mF\ll \tau_{ee}^{-1}$.
This crossover is shown by the curves (from red to violet) in Fig.~\ref{Fig6:I_ef}(b) with the zoom near the plateau $\langle I \rangle= e f$
in Fig.~\ref{Fig6:I_ef}(a). The parameters used are mentioned in the caption.
The representative curves in the quasiequilibrium regime
with the parameters $\delta = 10^{-2}\Delta$ and $l_{\rm mfp}=2.4\cdot 10^{-2}$ corresponding to $E_{\rm Th}/\Delta=1.8$ are shown by
dash-dotted green line in Fig.~\ref{Fig5:I_ef_3limits} and by dashed thick violet and red thin solid lines in panels (a), (b) and (c), (d) of Fig.~\ref{Fig6:I_ef}, respectively.

To go further to the equilibrium regime, $\tau_{eph}^{-1}\gg \Gamma_\mF$, one has to decrease the mean level spacing governing
both ratios \eqref{Gteph+Gtee}.
To keep electron-electron relaxation fixed in Fig.~\ref{Fig6:I_ef}(c), (d) we choose $E_{\rm Th}/\Delta=1.8$ and
$l_{\rm mfp}\propto (\delta/\Delta)^{-1/3}$.
We decrease $\delta/\Delta$ down to $10^{-7}$ where $l_{\rm mfp}$ achieves its maximal value equal to $1$.
However, as one can see from the solid blue curves in Fig.~\ref{Fig5:I_ef_3limits} and in panels (c)--(f) of Fig.~\ref{Fig6:I_ef},
even in this regime there is a small step in the plateau close to
$A_g=\Delta$ related to the non-equilibrium contribution to the turnstile current.

Finally, we consider the most nontrivial crossover from full non-equilibrium to equilibrium
which can be achieved by decreasing the parameter $\delta/\Delta$ from the maximal considered value $10^{-2}$ to $10^{-7}$ [see Fig.~\ref{Fig6:I_ef}(e), (f)].
In this case the Thouless energy $E_{\rm Th}\propto (\delta/\Delta)^{1/3}$ also changes
due to the fixed value of $l_{\rm mfp}=1$ needed to suppress electron-electron relaxation during the crossover.
Because of this, on some curves close to the non-equilibrium regime (red, orange, green) in the zoom Fig.~\ref{Fig6:I_ef}(e)
one can see an overshooting more typical for the quasiequilibrium regime.

It is important that the relaxation times $\tau_{ee},\tau_{eph}$, which determine $\mF(\ep)$
in Eq.~(\ref{f_stat_sol}), depend on the effective electronic temperature, $T_e$, via Eqs.~(\ref{SIA=}),~(\ref{taueph=}), while
$T_e$~itself is determined by $\mF(\ep)$ via Eq.~(\ref{T_e}), so $T_e$~has to be found self-consistently.
As a result, $T_e$~increases with~$A_g$, and so do the rates $\tau_{ee}^{-1},\tau_{eph}^{-1}$.
Thus, when~$A_g$ is changed, the device can switch between different regimes.
For example, the green curves in Figs.~\ref{Fig6:I_ef}(c)--(f) corresponding to $\delta/\Delta=10^{-4}$
go from quasiequilibrium (c), (d) or fully-nonequilibrium (e), (f) behavior at small $(A_g-A_g^{\rm ft})/\Delta$ to the equilibrium one at large amplitudes.

\section{Slow drive}
\label{App:slow_drive}
Now we consider the slow drive case, $f\ll \Gamma_{\mF}$, where $\Gamma_\mF$ is time-dependent and varies between $\gamma$ and $\gamma P_\infty$
during each half-period. Thus, we are assuming $f\ll \gamma P_\infty$.
Anticipating the results to be derived below, we note that in this case,
the leakage current turns out to be always large compared to $ef$, so
the SINIS SET at slow drive cannot be considered as a turnstile anymore.

Using the same assumptions as for the fast drive, $\mF_0=\mF_1\equiv \mF(\ep,t)$ and $\Gamma\gg \Gamma_{\mF},\tau_{ee}^{-1},\tau_{eph}^{-1}$, one can write the solutions \eqref{p_n_sol} for the occupation probabilities $p_n(t)$ and substitute them into Eq.~\eqref{kinetic_eq}, which can be written in the form of Eq.~\eqref{kinetic_eq_aver_f_nu},
\begin{align}
\dot{\mF}(\ep) = {}&{}
\Gamma_{\mF}(\ep,t)\left[\mF_{\rm neq}(\ep,t) - \mF(\ep)\right]
+{}\nonumber\\ {}&{} + \frac{\mF_{T_e(t)}(\ep) - \mF(\ep)}{\tau_{ee}}
+ \frac{\mF_{T}(\ep)-\mF(\ep)}{\tau_{eph}},
\label{App_eq:kinetic_eq}
\end{align}
with the time-dependent functions
\begin{align}
\Gamma_{\mF}(\ep,t) = {}&{}\gamma\sum_{\eta=\pm}[1-p_1(t)]\,
n_S^+\!\left(-\mu(t)+\eta \tfrac{e V}2-\ep\right) +{}\nonumber\\
{}&{}+ \gamma\sum_{\eta=\pm}p_1(t)\,
n_S^+\!\left(\mu(t)-\eta \tfrac{e V}2+\ep\right),\label{App_eq:Gamma_f(t)}\\
\label{App_eq:f_nu(t)}
\mF_{\rm neq}(\ep,t) = {}&{}\gamma\frac{1-p_1(t)}{\Gamma_{\mF}(\ep,t)}\sum_{\eta=\pm}n_S^+\left(-\mu(t)+\eta \tfrac{e V}2-\ep\right).
\end{align}
Here $\mu(t)$ is given by \eqref{E_g}. Note that $\tau_{ee},\tau_{eph}$ are also time-dependent, as they depend on the electronic temperature $T_e(t)$.
Due to the condition $\Gamma\gg\Gamma_\mF\gg{f}$, leakage
will be dominated by the longest part of each half-period
when $p_1(t)$ has reached its stationary value
($1-P_\infty$ and $P_\infty$ on the injection and the
ejection stage, respectively). Thus, we neglect all terms,
proportional to $e^{-\Gamma{t}}$.

\begin{figure}[h]
\center{
\includegraphics[width=0.40\textwidth]{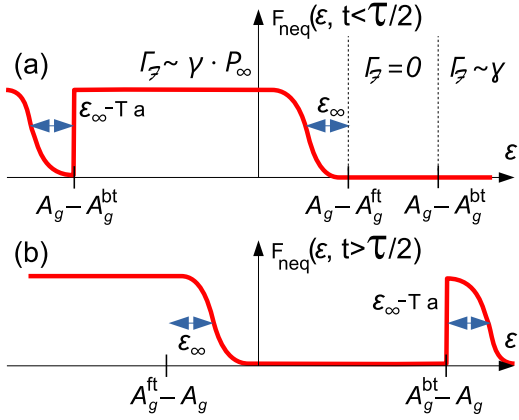}
}
\caption{(Color online)
A sketch of $\mF_{\rm neq}(\ep)$ in the
turnstile regime $A_g^{\rm ft}<A_g<A_g^{\rm bt}$ for a slow drive out of equilibrium $f, \tau_{eph}^{-1}, \tau_{ee}^{-1}\ll \Gamma_\mF$
at
(a)~the first and (b)~second halves of the period.
In panel (a) the approximate values of the distribution function relaxation rate $\Gamma_\mF$ due to tunneling are shown for the energy intervals (i-iii) mentioned in the text.}
\label{Fig8:Fneq_slow_drive}
\end{figure}

As we saw in Sec.~\ref{Sec:Fast_drive}, the most interesting is the non-equilibrium regime, $\Gamma_\mF\gg f, \tau_{eph}^{-1}, \tau_{ee}^{-1}$,
when the relaxation of the distribution function $\mF(\ep)$ is dominated by tunneling.
For the fast drive, turnstile errors were small in this regime; now we will show that for the slow drive this is no longer the case.
Due to the symmetry of the drive, $\mu(t+\mT/2) = -\mu(t)$, and of the resulting distribution function, $\mF(\ep,t+\mT/2) = 1-\mF(-\ep,t)$,
we consider the relaxation of $\mF(\ep,t)$ to its stationary value only in the first half-period $0<t<\mT/2$ with $\mu(0<t<\mT/2)=-A_g$.
One can identify three energy intervals:
(I)~$\ep>A_g+A_g^{\rm ft}$ where $\mF(\ep)$ relaxes to a small value $\mF_{\rm neq}\lesssim P_\infty e^{-\Delta/T}$
with the rate $\Gamma_{\mF}\sim \gamma$;
(II)~$A_g-A_g^{\rm ft}<\ep<A_g+A_g^{\rm ft}$ where $\Gamma_{\mF}=0$, the distribution function does not relax, and its value is determined by the relaxation during the other half-period;
(III)~$\ep<A_g-A_g^{\rm ft}$ where $\mF(\ep)$ relaxes with the rate $\Gamma_{\mF}\sim \gamma P_\infty$ to some nontrivial function considered below
[see Fig.~\ref{Fig8:Fneq_slow_drive}(a)].
Note that unlike the fast drive case,
we need a more accurate expression for the function
\begin{align}
n_S^+(\ep) ={}&{} n_S\left(\ep\right)\left[1-\mF_T\left(\ep\right)\right]
\approx\nonumber\\
\approx{}&{} n_S\left(\ep\right)\left[\theta\left(\ep-\Delta\right) + \theta\left(-\Delta-\ep\right)e^{\ep/T}\right],\label{App_eq:n_S^+}
\end{align}
including terms of order of $P_\infty\lesssim e^{-\Delta/T}$.

In the first interval, we denote $y=\ep-(A_g+A_g^{\rm ft})>0$ and obtain
\begin{align}
\Gamma_{\mF}(\ep) = {}&{}
\frac\gamma{2}\,n_S(\Delta+y)\left[1 + P_\infty e^{-(\Delta+y)/T}\right]+{}\nonumber \\
{}&{} + \frac\gamma{2}\,\theta(\Delta-eV+y)\, n_S(\Delta-eV+y)\times{}\nonumber\\
{}&{}\quad{}\times\left[1 + P_\infty e^{-(\Delta-eV+y)/T}\right] \sim \gamma,
\label{App_eq:Gamma_infty_E>A_g_+}
\end{align}
while
\begin{align}
\Gamma_{\mF}(\ep)\,\mF_{\rm neq}(\ep) = {}&{}
\frac\gamma{2}\,n_S(\Delta+y)P_\infty e^{-(\Delta+y)/T} +{}\nonumber\\
{}&{} + \frac\gamma{2}\,\theta(\Delta-eV+y)\, n_S(\Delta-eV+y)
\times{}\nonumber\\
{}&{}\quad{}\times P_\infty e^{-(\Delta-eV+y)/T} \lesssim
\gamma e^{-2\Delta/T}.\label{App_eq:Gamma_F_infty_E>A_g_+}
\end{align}
As we neglected the contributions of the order of $e^{-2\Delta/T}$ in \eqref{App_eq:n_S^+}, we should set
$\mF_{\rm neq}(\ep>A_g+A_g^{\rm ft})=0$ within this approximation in order to be consistent.

\begin{figure}[t]
\center{
\includegraphics[width=0.30\textwidth]{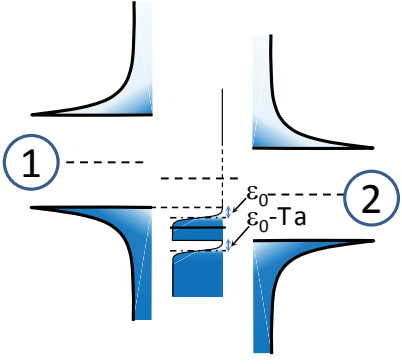}
}
\caption{(Color online)
The energy diagram of a SINIS SET turnstile in the regimes of slow drive out of equilibrium $f,\tau_{eph}^{-1}\ll \Gamma_{\mF}$.
The distribution function in the island shown by blue color corresponds to the first half-period.
The corresponding chemical potential is shown by the black solid line.
The shifts of the effective chemical potentials of the Fermi functions relatively to the edges of the superconducting DOSes
of the leads are shown explicitly.
}\label{Fig9:Slow_drive}
\end{figure}

In the third interval, we denote $x=A_g-A_g^{\rm ft}-\ep>0$
and obtain that the rate
\begin{align}
\Gamma_{\mF}(\ep) =  {}&{}
\frac\gamma{2}\,n_S(\Delta+x)\left[P_\infty  + e^{-(\Delta+x)/T}\right] +{}\nonumber\\
{}&{} + \frac\gamma{2}\,\theta(\Delta-eV+x)\, n_S(\Delta-eV+x)
\times{}\nonumber\\ {}&{}\quad\times
\left[P_\infty + e^{-(\Delta-eV+x)/T}\right] \sim\nonumber\\
\sim{}&{}\gamma \max\{P_\infty,e^{-\Delta/T}\},\label{App_eq:Gamma_infty_E<A_g_-}
\end{align}
and its product with the distribution function,
\begin{align}
\Gamma_{\mF}(\ep)\,\mF_{\rm neq}(\ep) = {}&{}
\frac\gamma{2}\,n_S(\Delta+x)\,P_\infty + {}\nonumber\\
{}&{}+\frac\gamma{2}\,\theta(\Delta-eV+x)\, n_S(\Delta-eV+x)P_\infty\nonumber\\
\sim {}&{}\Gamma_{\mF}(\ep),\label{App_eq:Gamma_F_infty_E<A_g_-}
\end{align}
are of the same order.
Parameterizing $P_\infty = e^{-(\Delta+\ep_\infty)/T}$ by a positive
energy $\ep_\infty>0$, we can write the resulting distribution function in
the form of a double Fermi function (see Figs.~\ref{Fig8:Fneq_slow_drive},~\ref{Fig9:Slow_drive})
\begin{gather}\label{App_eq:f(E,t)_slow_drive}
\mF_{\rm neq}(\ep) = \left\{\begin{array}{ll}
\mF_T(\ep_\infty-x), & x<e V,\\
\mF_T(\ep_\infty + e V - x - T a), & x > e V.
\end{array}\right.
\end{gather}
Here $a = \ln[1+n_S(\Delta + x)/n_S(\Delta - e V + x)]$ is bounded by $1<a<2$, and we neglected exponentially small factors
$e^{-(e V - \ep_\infty)/T}$ for $e V - \ep_\infty\gg T>0$.
For the case $\ep_\infty-e V\gg T>0$ function \eqref{App_eq:f(E,t)_slow_drive} becomes the Fermi-Dirac function
$\mF_{\rm neq}(\ep)=\mF_T(\ep_\infty + e V - x - T a)$ with zero values at $x < e V$.


By definition, $P_\infty$~is determined by the distribution function itself via Eqs.~(\ref{P_infty}), (\ref{w+=}), (\ref{w-=}).
Only the values of $\mF(\ep)$ in the intervals (I)~and~(III) enter the integrals in Eqs.~(\ref{w+=}), (\ref{w-=}),
and when one substitutes there $\mF(\ep)$ found above, Eq.~(\ref{P_infty}) becomes an identity.
Thus, to find $P_\infty$ or $\ep_\infty$, it is necessary to use constraint \eqref{F_n_fixed_mu},
\begin{gather}\label{App_eq:F_fixed_mu_condition}
\frac{2}{\delta}\int[\mF(\ep)-\theta(-\ep)]\,d\ep=p_1.
\end{gather}
This equation is sensitive to $\mF(\ep)$ in the nonrelaxing interval~(II),
determined by the relaxation on the previous half-period. It is important that for $A_g>A_g^{\rm ft}$, interval~(II) maps
on interval~(III) of the previous half-period, so there is no uncertainty in $\mF(\ep)$.
The solution for $\ep_\infty$ depends on the relation between $A_g$, $eV$,
and $\Delta$, and the allowed region $A_g>\Delta-eV/2$, $0<eV<2\Delta$ of the $(eV,A_g)$ plane splits into $7$ subregions.
We study in detail the simplest case of $\mF(A_g-A_g^{\rm ft}<\ep<A_g+A_g^{\rm ft}) = 0$,
realized when either $\Delta+eV/6<A_g<\Delta+3eV/2$ or $\Delta-eV/2<A_g<eV/2$.
For other arrangements of $A_g$, $eV$, and $\Delta$, the distribution has a different shape, but the results for the leakage current
are qualitatively similar to the one obtained below, so we do not give details for these cases.

In the simplest case of $\mF(A_g-A_g^{\rm ft}<\ep<A_g+A_g^{\rm ft}) = 0$, we obtain $\ep_\infty \approx (A_g - A_g^{\rm ft})/2$
which is just the conservation of the chemical potential (the number of filled states,
$(A_g - A_g^{\rm ft}-\ep_\infty)/\delta$, above the Fermi energy should be equal to the number of empty states, $\sim\ep_\infty/\delta$, below).
The corresponding distribution
is shown in Figs.~\ref{Fig8:Fneq_slow_drive} and \ref{Fig9:Slow_drive}.
The relaxation bottleneck is the discharging process with the rates $\Gamma_{j-}\sim \gamma n_{qp}e^{-(|\ep|-\Delta-eV_j)}[1-\mF(\ep)]$
occurring in the first half-period due to a small density of hole-like quasiparticles $n_{qp}\sim e^{-\Delta/T}$ with a narrow thermal distribution,
$e^{-(|\ep|-\Delta-eV_j)}$, concentrated near the quasiparticle band edge, $-\Delta-eV_j$.
The charging process is suppressed by the factor $p_0(t)=P_\infty$ of the small probability for the island to be discharged.
The balance between the charging rate, $P_\infty \Gamma_{1+}$, with a weak energy dependence, and the discharging rates $\Gamma_{j-}$
in both electrodes with the thermal exponential energy dependence $e^{-(|\ep|-\Delta-eV_j)}$ formes the double Fermi distribution $\mF(\ep)$.
The effective chemical potentials of these Fermi functions are shifted from the gap edges to nearly the same value $\simeq\ep_\infty$
which conserves the overall chemical potential of $\mF_{\rm neq}$ and determines $P_\infty$ via Eq.~\eqref{App_eq:F_fixed_mu_condition}.

It is quite easy to see now that the described relaxation dynamics
leads to a large leakage current. Indeed, using the symmetry
$\mF(\ep,t+\mT/2) = 1-\mF(-\ep,t)$, one can see that the
distribution function changes significantly from one stationary
state, $\mF_{\rm neq}(\ep)$ [Eq.~\eqref{App_eq:f(E,t)_slow_drive}]
on the first half-period, to $1-\mF_{\rm neq}(-\ep)$ on the second
one (Fig.~\ref{Fig8:Fneq_slow_drive}). To produce this change, a large
number of electrons, $\sim(A_g-A_g^{\rm ft})/\delta$, should
tunnel out from (into) the island during the relaxation stage
($0<t\lesssim\Gamma_\mF^{-1}$ and
$\mT/2<t\lesssim\mT/2+\Gamma_\mF^{-1}$ on the two half-periods).
This gives a contribution to the leakage current
\begin{gather}\label{App_eq:I_leak_relax}
\frac{I_{\rm leak}}{e f}\sim \frac{A_g-A_g^{\rm ft}}{\delta}\gg 1.
\end{gather}
Moreover, the long interval $\Gamma_\mF^{-1}\lesssim t<\mT/2$,
when $\mF(\ep)$ has already relaxed, contributes even more to
the leakage current. Indeed, using the expression \eqref{I_leak}
for $I_{\rm leak}$, the estimates
$\Gamma_{1-}\sim\Gamma_{2-}\sim \Gamma P_\infty$ from Eq.~\eqref{P_infty}, and the estimate $\Gamma\sim\Gamma_{1+}\sim \gamma \ep_\infty/\delta$,
proportional to the number of empty states in the island below the gaps in the electrodes, we estimate the second contribution to the leakage current as
\begin{gather}\label{App_eq:I_leak_stationary}
I_{\rm leak} \sim e \Gamma_{2-} \sim e \gamma P_\infty \frac{\ep_\infty}{\delta}\gg e f.
\end{gather}
The latter inequality is governed by two large parameters $\ep_\infty/\delta\sim (A_g-A_g^{\rm ft})/\delta$ and $\gamma P_\infty/f$.

Among other regions in the $(A_g,eV)$ plane, not corresponding to $\mF(A_g-A_g^{\rm ft}<\ep<A_g+A_g^{\rm ft}) = 0$, we mention two more, defined by
either $\max\{\Delta-eV/2,3eV/2-\Delta\}<A_g<\Delta$ with $\ep_\infty = A_g - A_g^{\rm ft}$
or $A_g>\Delta+3eV/2$ with $\ep_\infty = A_g - \Delta - e V/2>e V$.
In both these cases the distribution $\mF_{\rm neq}(\ep)$ coincides on the two half-periods,
$\mF_{\rm neq}(\ep,t<\mT/2)=\mF_{\rm neq}(\ep,t>\mT/2)$ being the Fermi-Dirac one in the latter one.
Then, the contribution \eqref{App_eq:I_leak_relax} to the leakage, related to the relaxation of $\mF(\ep)$, vanishes.
Still, the stationary leakage current estimated by \eqref{App_eq:I_leak_stationary} remains large.


Let us now briefly discuss the quasiequilibrium regime, $\tau_{ee}^{-1}\gg \Gamma_{\mF}\gg f,\tau_{eph}^{-1}$, when the distribution function deviates slightly from $\mF_{T_e(t)}(\ep)$. Generally speaking, one has to take into account the time dependence of the effective electronic temperature $T_e(t)$ governed by \eqref{T_e}.
However, for the symmetric drive, Eq.~(\ref{E_g}), and the symmetric device, $C_1=C_2$, $G_1=G_2$, $\gamma_1=\gamma_2=\gamma$,
$-V_1=V_2=V/2>0$, the steady state values of the electronic temperature in both halves of the period are equal.
Variations of $T_e(t)$ in this limit occur only in the short charging/discharging time intervals $\sim 1/\Gamma$ and can be neglected.
Then, the treatment of the quasiequilibrium regime is quite analogous to Sec.~\ref{Sec:Fast_ee_relax}, with an analogous result (large leakage).

In equilibrium, $\tau_{eph}^{-1}\gg \Gamma_{\mF}\gg f$, the electron distribution  is Fermi-Dirac. The estimate for the leakage current is then the same as in
Eq.~(\ref{App_eq:I_leak_stationary}), with the replacement $\ep_\infty\to{A}_g-A_g^{\rm ft}$, again giving large leakage.

\section{Conclusion}

We have shown that the accuracy of the pumping current quantization in the turnstile regime of the hybrid SINIS SET is a non-monotonic function of 
the relaxation rate of the electronic distribution function on the island due to tunneling, $\Gamma_{\mF}\sim \gamma \left[P_\infty+f /\Gamma\right]$.
In the equilibrium regime, $\Gamma_{\mF}\gg\tau_{eph}^{-1}$, the turnstile current has a plateau $\langle I\rangle \simeq e f$ in the standard interval of driving amplitudes, $A_g^{\rm ft}<A_g<A_g^{\rm bt}$ with the forward and backward tunneling thresholds $A_g^{\rm ft,bt}=\Delta\mp{e}V/2$.
Increase of the tunneling relaxation rate $\Gamma_{\mF}$ brings the system into quasiequilibrium at $\tau_{eph}^{-1}\ll \Gamma_{\mF}\ll \tau_{ee}^{-1}$, with the effective electronic temperature strongly different from the bath temperature. In this regime, turnstile errors are large except in the small interval near the threshold $A_g/A_g^{\rm ft}-1\ll 1$.
Surprisingly, at even faster tunneling relaxation rate $\Gamma_{\mF}\gg\tau_{ee}^{-1},\tau_{eph}^{-1}$, when the electronic distribution on the island is essentially non-thermal, the turnstile plateau is recovered, but in a smaller interval of driving amplitudes, $A_g^{\rm ft}<A_g<\Delta$.
These considerations hold for sufficiently fast driving frequency, $\Gamma_{\mF}\ll f$; in the opposite case, the tunneling processes change the distribution function significantly over the period and lead to a large leakage current.

Such a non-monotonic behavior of the current plateau indicates that the turnstile operation is mostly spoiled by the electron-electron relaxation forming the long-tailed distribution function, but not by the driving itself.

\acknowledgements
We would like to thank J. P. Pekola 
for useful discussions.
This work has been supported in part
by the Nanosciences Foundation under the aegis of the Joseph Fourier University Foundation (Grenoble, France),
by the Russian Foundation for Basic Research, and the grant of the Russian Science Foundation (No. 15-12-10020).

\appendix

%
%
%


\section{Verification of the assumption $\mF_0(\ep,t)=\mF_1(\ep,t)$}\label{App:check_f_0=f_1}
To verify the validity of the assumption that the $n$-dependence of the distribution functions $\mF_n(\ep)$ is negligible,
one has to start with the rate equations for the full density matrix, see Eqs.~(3) in Ref.~\onlinecite{Averin-Korotkov_JETP90}.
For $n=0,1$ within the assumption of the energy scale separation $E_C\gg\delta$, one can rewrite these equations in the form of Eq.~\eqref{rate_eq_pn}
and the kinetic equations for the distribution functions $\mF_0(\ep)$ and $\mF_1(\ep)$:
\begin{eqnarray}
\label{App_eq:kinetic_eq_f_0}
p_0 \dot{\mF}_0 &= {}&{} \Gamma_{-} p_1 (\mF_1-\mF_0)\nonumber\\
{}&{}+&\gamma{p}_0(t)\sum_{\eta=\pm}
n_S^+\left(-\mu(t)+\eta \tfrac{e V}2-\ep\right)\mF_0(1-\mF_0) \nonumber\\
{}&{}+& \gamma{p}_1(t)\sum_{\eta=\pm}n_S^+\left(\mu(t)-\eta \tfrac{e V}2+\ep\right)\mF_1^2,\\
\label{App_eq:kinetic_eq_f_1}
p_1 \dot{\mF}_1 &=& \Gamma_{+} p_0 (\mF_0-\mF_1)\nonumber\\
{}&{}+& \gamma{p}_0(t)\sum_{\eta=\pm}n_S^+\left(-\mu(t)+\eta \tfrac{e V}2-\ep\right)(1-\mF_0)^2 \nonumber\\
{}&{}+& \gamma{p}_1(t)\sum_{\eta=\pm}n_S^+\left(\mu(t)-\eta \tfrac{e V}2+\ep\right)\mF_1(1-\mF_1),
\end{eqnarray}
where $\Gamma_\pm = \sum_j \Gamma_{j\pm}$.
As we will show below, the relaxation time to the state $\mF_0(\ep,t)=\mF_1(\ep,t)\equiv \mF(\ep,t)$ is of the same order as the charge relaxation time $\Gamma^{-1}$ which is much shorter than all other characteristic times.

For this purpose let us consider the beginning of the injection stage.
Just before this stage, the island has been discharged, therefore $p_1(0)\simeq P_\infty\ll 1$, while $p_0(0)\simeq 1$.
As we suddenly changed $\mu(t)$ from $-A_g$ to $A_g$ we made $\Gamma_{+}\simeq \Gamma\gg \Gamma_{-},\gamma$.
As a result, all the rates in \eqref{App_eq:kinetic_eq_f_0} are much smaller than $\Gamma$, so we can consider $\mF_0(\ep,t)=\bar{\mF_0}(\ep)$ to be constant at this time scale.
On the other hand, the first term on the right-hand side of Eq.~\eqref{App_eq:kinetic_eq_f_1} is dominant at such short times, and one can rewrite this equation as follows:
\begin{gather}\label{App_eq:kinetic_eq_p_1_f_1}
\frac{d(p_1 \mF_1)}{dt} = \Gamma_{+} p_0 \bar{ \mF_0} - \Gamma_{-} \cdot(p_1 \mF_1)
 \ ,
\end{gather}
where we used the equation
\begin{gather}
\frac{d p_0}{dt} = -\Gamma_{+} p_0 + \Gamma_{-} p_1
\end{gather}
with the solution \eqref{p_n_sol_1st_half} $p_0(t)\simeq P_\infty+ (1-2P_\infty)e^{-\Gamma t}$, and neglected an exponentially small correction $e^{-\Gamma\mT/2}$.

Finally, integrating \eqref{App_eq:kinetic_eq_p_1_f_1} at $t\lesssim \Gamma^{-1}$, one can find
\begin{gather}
\mF_1(\ep,t)=\bar{\mF_0} +\frac{P_\infty(\mF_1(0)-\bar{\mF_0})e^{-\Gamma_{-} t}}{P_\infty +(1-2P_\infty) (1-e^{-\Gamma t})} \ .
\end{gather}
One can see that the function $\mF_1(\ep,t)$ relaxes from any initial value $\mF_1(0)$ to the final one $\simeq\bar{\mF_0}+P_\infty[\mF_1(0)-\bar{\mF_0}]$, which is close to $\bar{\mF_0}$, on the time scale $\Gamma^{-1}$ with a subsequent relaxation at larger time scales.

Eventually neglecting such small difference $P_\infty=\Gamma_{-}/\Gamma\ll 1$ and working at larger time scales $\Gamma t\gtrsim 1$ one can put $\mF_0(\ep,t)=\mF_1(\ep,t)\equiv \mF(\ep,t)$.


\begin{thebibliography} {99}
\bibitem{Giazotto2006}%
F. Giazotto, T. T. Heikkil\"a, A. Luukanen, A. M. Savin, and J. P. Pekola,
Opportunities for mesoscopics in thermometry and refrigeration: Physics and applications.
{Rev.~Mod.~Phys.} {\bf 78}, 217 (2006).

\bibitem{Pekola_RMP_2013}
J.~P.~Pekola, O.-P.~Saira, V.~F.~Maisi, A.~Kemppinen, M.~M{\" o}tt{\" o}nen, Y.~A.~Pashkin, and D.~V.~Averin
Rev. Mod. Phys. {\bf 85}, 1421 (2013).

\bibitem {Knowles2012}%
H.~S.~Knowles, V.~F.~Maisi, and J.~P.~Pekola, 
Appl. Phys.  Lett. {\bf100}, 262601 (2012).

\bibitem{Footnote_QP_poisoning}
For example, there are such effects as decrease of the quality factors of S resonators \cite{Wang2009,Barends2011},
decoherence in qubit systems \cite{Martinis2009,Paik2011,Corcoles2011},
the excess current in single-electron turnstiles \cite{Knowles2012},
and low efficiency of electronic cooling in normal metal (N) - insulator (I) - superconductor (S) junctions \cite{Pekola2000,Rajauria2009}.

\bibitem {Wang2009}
H. Wang, M. Hofheinz, J. Wenner, M. Ansmann, R. C. Bialczak, M. Lenander, E. Lucero, M. Neeley, A. D. O'Connell, D. Sank, M. Weides, A. N. Cleland, and J. M. Martinis,
Appl. Phys. Lett. {\bf 95}, 233508 (2009).

\bibitem {Barends2011}%
R. Barends, J. Wenner, M. Lenander, Y. Chen, R. C. Bialczak, J. Kelly, E. Lucero, P. O'Malley, M. Mariantoni, D. Sank, H. Wang, T. C. White, Y. Yin, J. Zhao, A. N. Cleland, J. M. Martinis, and J. J. A. Baselmans,
Appl. Phys. Lett. {\bf 99}, 113507 (2011).

\bibitem {Martinis2009}	
J. M. Martinis, M. Ansmann, and J. Aumentado,
Phys. Rev. Lett. {\bf 103}, 097002 (2009).

\bibitem {Paik2011}
H. Paik, D. I. Schuster, L. S. Bishop, G. Kirchmair, G. Catelani, A. P. Sears, B. R. Johnson, M. J. Reagor, L. Frunzio, L. I. Glazman, S. M. Girvin, M. H. Devoret, and R. J. Schoelkopf,
Phys. Rev. Lett. {\bf 107}, 240501 (2011).

\bibitem {Corcoles2011}%
A. D. C\`orcoles, J. M. Chow, J. M. Gambetta, C. Rigetti, J. R. Rozen, G. A. Keefe, M. B. Rothwell, M. B. Ketchen, and M. Steffen,
Appl. Phys.  Lett. {\bf 99}, 181906 (2011).

\bibitem {Pekola2000}
J. P. Pekola, D. V. Anghel, T. I. Suppula, J. K. Suoknuuti, A. J. Manninen, and M. Manninen,
Appl. Phys. Lett. {\bf 76}, 2782 (2000).

\bibitem {Rajauria2009}
S. Rajauria, H. Courtois, and B. Pannetier,
Phys. Rev. B {\bf 80}, 214521 (2009).
\bibitem {Peltonen2011}
J. T. Peltonen, J. T. Muhonen, M. Meschke, N. B. Kopnin, and J. P. Pekola,
Phys. Rev. B {\bf 84}, 220502(R) (2011).

\bibitem {Nsanzineza2014}
I. Nsanzineza and B. L. T. Plourde,
Phys. Rev. Lett. {\bf 113}, 117002 (2014).

\bibitem {Wang2014}
C. Wang, Y. Y. Gao, I. M. Pop, U. Vool, C. Axline, T. Brecht, R. W. Heeres, L. Frunzio, M. H. Devoret,
G. Catelani, L. I. Glazman, and R. J. Schoelkopf,
Nature Commun. {\bf 5}, 5836 (2014).

\bibitem{Vool2014}
U. Vool, I. M. Pop, K. Sliwa, B. Abdo, C. Wang, T. Brecht, Y. Y. Gao, S. Shankar, M. Hatridge, G. Catelani,
M. Mirrahimi, L. Frunzio, R. J. Schoelkopf, L. I. Glazman, and M. H. Devoret,
Phys. Rev. Lett. {\bf 113}, 247001 (2014).

\bibitem {Woerkom2015}
D. J. van Woerkom, A. Geresdi, and L. P. Kouwenhoven,
Nature Phys., {\bf 11}, 547-550 (2015).

\bibitem{Taupin2016_NatComm}
M. Taupin, I. M. Khaymovich, M. Meschke, A. S. Mel'nikov, and J. P. Pekola,
Nature Commun. {\bf 7}, 10977 (2016).
\bibitem{Blamire1991}
M. G. Blamire, E. C. G. Kirk, J. E. Evetts, and T. M. Klapwijk,
Phys. Rev. Lett. {\bf 66}, 220-223 (1991).

\bibitem{Heslinga1993}
D. R. Heslinga and T. M. Klapwijk,
Phys. Rev. B {\bf 47}, 5157-5164 (1993).

\bibitem {Goldie1990}%
D. J. Goldie, N. E. Booth, C. Patel, and G. L. Salmon,
Phys. Rev. Lett. {\bf 64}, 954-957 (1990).

\bibitem {Nguyen2013}	
H. Q. Nguyen, T. Aref, V. J. Kauppila, M. Meschke,
C. B. Winkelmann, H. Courtois, and J. P. Pekola,
New J. Phys. {\bf 15}, 085013 (2013).
\bibitem{Bespalov_Nazarov_2016}
A. Bespalov, M. Houzet, J. S. Meyer, and Y. V. Nazarov,
arXiv:1603.04273.

\bibitem{Pekola_PRL_2010}
J. P. Pekola, V. F. Maisi, S. Kafanov, N. Chekurov, A. Kemppinen, Yu.~A. Pashkin, O.-P. Saira,
M. M{\"o}tt{\"o}nen, and J. S. Tsai,
Phys.~Rev.~Lett. {\bf 105}, 026803 (2010).

\bibitem{Saira2012}
O.-P. Saira, A. Kemppinen, V. F. Maisi, and J. P. Pekola,
Phys.~Rev.~B {\bf 85}, 012504 (2012).

\bibitem{Riste2013}
D. Riste, C. C. Bultink, M. J. Tiggelman, R. N. Schouten, K. W. Lehnert, and L. DiCarlo,
Nature Comm. {\bf 4}: 1913 (2013).
\bibitem{Pothier1997}
H. Pothier, S. Gueron, N. O. Birge, D. Esteve, and M. H. Devoret,
Phys. Rev. Lett.~{\bf 79}, 3490 (1997); {\it ibid} Z. Phys. B {\bf 104}, 178 (1997).

\bibitem{Nazarov_Heikkila2010}
M.~A.~Laakso, T.~T.~Heikkil{\"a} and Y.~V.~Nazarov,
Phys. Rev. Lett. 104, 196805 (2010)

\bibitem{Pekola_Heikkila_Non-eq_f(E)-1}
J. P. Pekola, T. T. Heikkil{\"a}, A. M. Savin, J. T. Flyktman, F. Giazotto, F. W. J. Hekking,
Phys. Rev. Lett. {\bf 92}, 056804 (2004).

\bibitem{Pekola_Heikkila_Non-eq_f(E)-2}
F. Giazotto, T. T. Heikkil{\"a}, F. Taddei, Rosario Fazio, J. P. Pekola, F. Beltram,
Phys. Rev. Lett. {\bf 92}, 137001 (2004).

\bibitem{SINIS_turnstile_Nat2008}
J.~P.~Pekola, J.~J.~Vartiainen, M. M{\"o}tt{\"o}nen, O.-P.~Saira, M.~Meschke, and D.~V.~Averin,
Nature Physics {\bf 4}, 120 (2008).

\bibitem{Kashcheyevs_RPP2015}
B.~Kaestner and V.~Kashcheyevs,
Rep. Prog. Phys. 78, 103901 (2015)

\bibitem{vanZanten2016}
D.~M.~T.~van Zanten, D.~M.~Basko, I.~M.~Khaymovich, J.~P.~Pekola, H.~Courtois, C.~B.~Winkelmann,
Phys. Rev. Lett. {\bf116}, 166801 (2016).

\bibitem{Aleiner2002}
I.~L.~Aleiner, P.~W.~Brouwer, and L.~I.~Glazman,
Phys. Rep., {\bf358}, 309 (2002).

\bibitem{Averin-Korotkov_JETP90}
D. V. Averin, A. N. Korotkov,
JETP {\bf 70}, 937 (1990) [Zh. Eksp. Teor. Fiz. {\bf 97}, 1661 (1990)].

\bibitem{Averin-Korotkov_JLTP90}
D. V. Averin, A. N. Korotkov,
J. Low. Temp. Phys. {\bf 80}, 173 (1990).

\bibitem{Footnote_mu_in_St}
Strictly speaking, since both electron-electron and electron-phonon
collisions conserve the total number of electrons on the island,
both limiting equilibrium distributions should be taken with some
chemical potential, chosen to give the same total number of
electrons as the distribution~$\mF(\ep)$. However, in this work
we consider only two values of this total number, so the chemical
potential may vary only by a small amount $\sim\delta$, which
we neglect.

\bibitem{Sivan1994}
U. Sivan, Y. Imry, and A. G. Aronov,
Europhys. Lett. {\bf 28}, 115 (1994).

\bibitem{Blanter1996}
Ya. M. Blanter,
Phys. Rev.~B {\bf 54}, 12807 (1996).


\bibitem{Wellstood1994}
F. C. Wellstood, C. Urbina, and J. Clarke,
Phys. Rev. B {\bf 49}, 5942 (1994).

\bibitem{Sergeev2000}
A. Sergeev and V. Mitin,
Phys. Rev. B {\bf 61}, 6041 (2000).

\bibitem{Yudson2003}
V. I. Yudson and V. E Kravtsov,
Phys. Rev. B {\bf 67}, 155310 (2003).

\bibitem{Basko2005}
D. M. Basko and V. E. Kravtsov,
Phys. Rev. B {\bf 71}, 085311 (2005).

\bibitem{Sergeev2005}
A. Sergeev, M. Yu. Reizer, and V. Mitin,
Phys. Rev. Lett. {\bf 94}, 136602 (2005).

\bibitem{Prunnila2005}
M. Prunnila, P. Kivinen, A. Savin, P. T\"orm\"a, and J. Ahopelto,
Phys. Rev. Lett. {\bf 95}, 206602 (2005).

\bibitem{AverinPekola2008}
D. V. Averin and J. P. Pekola,
Phys. Rev. Lett. {\bf 101}, 066801 (2008).

\bibitem{Footnote_QEq_small_rates}
Here we neglect the exponentials
$e^{-\Delta/T-c_0(A_g-\Delta\pm e V/2)/(A_g-A_g^{\rm ft})}$ due to the smallness of the bath temperature $T<T_e/2$.

\bibitem{Heikkila_SISIS_non-eq_f(E)}
M. A. Laakso, P. Virtanen, F. Giazotto, and T. T. Heikkil{\"a},
Phys. Rev. B {\bf 75}, 094507 (2007).

\bibitem{Footnote_AlMn}
Although aluminum itself is superconducting at low temperatures and cannot be used as a material for a normal island,
a small admixture of Mn around $0.1-0.3$~$\%$ suppresses the critical temperature of the AlMn alloy down to $50$~mK
(Refs.~\onlinecite{Ullom_JLTP2004,Ullom_PRL2004}).
Such a small amount of Mn impurities does not significantly change the thermal properties of the material such as heat capacity.\cite{Ullom_JLTP2004}
We also use the unchanged value of $\Sigma=0.2\cdot 10^{9}$~J~K$^{-5}$~m$^{-3}$ taken for Al, although there are some indirect indications
of larger $\Sigma$ values in this material (see, e.g., Ref.~\onlinecite{Ullom_Clark_APL2004_AlMn_fridge}), but
we are not aware of any direct measurements of the electron-phonon coupling constant in AlMn.

\bibitem{Ullom_JLTP2004}
S. T. Ruggiero, A. Williams, W. H. Rippard, A. Clark, S. W. Deiker, L. R. Vale, and J. N. Ullom,
J. Low Temp. Phys., {\bf134}, 973 (2004).

\bibitem{Ullom_PRL2004}
G. O�Neil, D. Schmidt, N. A. Miller, J. N. Ullom, A. Williams, G. B. Arnold, and S. T. Ruggiero,
Phys. Rev. Lett., {\bf100}, 056804 (2004).

\bibitem{Ullom_Clark_APL2004_AlMn_fridge}
A. M. Clark, A. Williams, S. T. Ruggiero, M. L. van den Berg, and J. N. Ullom,
Appl. Phys. Lett., {\bf 84}, 625 (2004).

\bibitem{SINIS_turnstile_high_EC}
A. Kemppinen, S. Kafanov, Yu. A. Pashkin, J. S. Tsai, D. V. Averin, and J. P. Pekola,
Appl.~Phys.~Lett. {\bf 94}, 172108 (2009).

\end{thebibliography}
\end{document}